\documentclass[a4paper,11pt]{article}
\usepackage{a4wide,amsmath,amssymb,bbm}
\usepackage[english]{babel}

\usepackage{color}

\usepackage[pdfencoding=auto,psdextra]{hyperref}
\usepackage{enumerate}
\hypersetup{
    colorlinks,
    linkcolor={black},
    citecolor={black},
    urlcolor={black}
}

\usepackage{mathrsfs}

\newcommand{\be}{\begin{equation}}
\newcommand{\ee}{\end{equation}}

\newcommand{\xt}{\underline{x}}
\newcommand{\eqcf}{F}

\makeatletter

\@addtoreset{equation}{section} \makeatother

\begin{document}


\vspace{30pt}

\begin{center}
{\huge{\bf Two-loop conformal invariance for Yang-Baxter deformed strings}}

\vspace{80pt}

Riccardo Borsato$\, ^{a}$ \ \ and \ \ Linus Wulff$\, ^b$

\vspace{15pt}

{
\small {$^a$\it 
Instituto Galego de F\'isica de Altas Enerx\'ias (IGFAE), Universidade de  Santiago de Compostela, Spain}\\
\vspace{5pt}
\small {$^b$\it Department of Theoretical Physics and Astrophysics, Masaryk University, 611 37 Brno, Czech Republic}
\\
\vspace{12pt}
\texttt{riccardo.borsato@usc.es, wulff@physics.muni.cz}}\\

\vspace{100pt}

{\bf Abstract}
\end{center}
\noindent
The so-called homogeneous Yang-Baxter (YB) deformations can be considered a non-abelian generalization of T-duality--shift--T-duality (TsT) transformations. TsT transformations are known to preserve conformal symmetry to all orders in $\alpha'$. Here we argue that (unimodular) YB deformations of a bosonic string also preserve conformal symmetry, at least to two--loop order. We do this by showing that, starting from a background with no NSNS-flux, the deformed background solves the $\alpha'$--corrected supergravity equations to second order in the deformation parameter. At the same time we determine the required $\alpha'$--corrections of the deformed background, which take a relatively simple form. In examples that can be constructed using, possibly non-commuting sequences of, TsT transformations we show how to obtain the first $\alpha'$--correction to all orders in the deformation parameter by making use of the $\alpha'$-corrected T-duality rules. We demonstrate this on the specific example of YB deformations of a Bianchi type II background.

\pagebreak 
\tableofcontents

\setcounter{page}{1}

\section{Introduction and summary of results}
Yang-Baxter (YB) deformations were first introduced by Klim\v cik in \cite{Klimcik:2002zj}. It was later understood that they have the remarkable property of preserving integrability \cite{Klimcik:2008eq}. If one starts from an integrable sigma model and performs a YB deformation the resulting model is also integrable. This made people interested in applying them in string theory, which was done for the $AdS_5\times S^5$ superstring in \cite{Delduc:2013qra,Kawaguchi:2014qwa}. The YB deformation is based on an R-matrix for which there are two basic possibilities-- $R$ can solve either the classical Yang-Baxter equation (CYBE) or the modified classical Yang-Baxter equation (mCYBE). The former case is often referred to as \emph{homogeneous} YB deformations and is the case we consider here. It was shown in \cite{Borsato:2016ose} that these models typically have a Weyl-anomaly\footnote{This manifests itself, in the superstring case, as a target space solving the generalized supergravity equations \cite{Arutyunov:2015mqj,Wulff:2016tju} rather than the standard ones.} unless the R-matrix is unimodular, i.e. its contraction with the structure constants of the isometry algebra of the original model vanishes $R^{IJ}f_{IJ}{}^K=0$. This is similar to the anomaly encountered in non-abelian T-duality (NATD) \cite{delaOssa:1992vci} on a non-unimodular group \cite{Gasperini:1993nz,Alvarez:1994np,Elitzur:1994ri}. Indeed it was argued in \cite{Hoare:2016wsk} that homogeneous YB deformations should have a realization in terms of NATD and this was then proven in \cite{Borsato:2016pas} (see also \cite{Borsato:2017qsx}). While the original YB deformations were defined only for sigma models of the symmetric space type, the realization of the homogeneous models using NATD meant that they could be defined for a general string sigma model with isometries. This was carried out for the Green-Schwarz superstring in \cite{Borsato:2018idb} and rules for writing the supergravity background directly in terms of the R-matrix were derived.\footnote{These rules were first guessed, at the supergravity level and restricted to the case of vanishing NSNS flux, in \cite{Araujo:2017jkb} (see also \cite{Bakhmatov:2017joy} and~\cite{Sakamoto:2017cpu}).}

The simplest class of such YB deformations is when $R$ is defined on an abelian subalgebra of the isometry algebra. In this case the deformation is equivalent to a T-duality--shift--T-duality (TsT) transformation \cite{Osten:2016dvf}. These are also known as $O(d,d)$-transformations \cite{Meissner:1991zj,Meissner:1991ge} and they have been argued to map a consistent string background to another consistent string background, i.e. there should exist corrections to the background fields such that the corrected background solves the $\alpha'$--corrected supergravity equations to all orders in $\alpha'$ \cite{Sen:1991zi,Gasperini:1991qy,Hassan:1992gi,Kiritsis:1992uz,Panvel:1992he,Kiritsis:1993ju}.\footnote{Note however that the form of the $\alpha'$--corrections are only known in special cases and to low loop order, e.g. \cite{Panvel:1992he}.} Here we want to ask what happens for YB deformations in general at the quantum level.\footnote{Homogeneous YB deformations also have an $O(d,d)$ interpretation as so called $\beta$-shifts \cite{Lust:2018jsx,Sakamoto:2018krs}.} Unimodular YB deformations are known to give a conformal theory at one loop, i.e. the background solves the (super)gravity equations. Here we will analyze the two-loop equations in the bosonic string case. For simplicity we will restrict to deformations of backgrounds with vanishing NSNS-flux. We will show, to second order in the deformation parameter, that the deformed background can be corrected so that it solves the 2--loop equations. Furthermore the correction to the background fields can be cast in a relatively simple form.
{Using the knowledge of the full corrections in special cases derived using T-duality (see below), we write an expression to all orders in the deformation parameter, which works in some simple cases but not in general.
}

Since the homogeneous YB deformations can be constructed using NATD, our results indicate that also NATD should preserve conformality at two loops, and possibly all orders in $\alpha'$. {A convincing argument for the preservation of conformality for NATD would follow from a generic analysis to all orders in the deformation parameter $\eta$, since NATD is recovered in a $\eta\to\infty$ limit.} Another piece of evidence for this comes from the recent analysis of renormalizability of deformed sigma models with two-dimensional target space in \cite{Hoare:2019ark}, and very recently~\cite{Hoare:2019mcc} {(see also \cite{Eghbali:2018ohx})}. Some of the deformations considered have a limit where they reduce to NATD and it was found that the models behave nicely beyond lowest order in $\alpha'$ suggesting that things should work out to all orders in $\alpha'$.

For YB deformations of TsT--type we can also exploit another method to obtain explicit $\alpha'$--corrections and to promote those backgrounds to two-loop solutions. We can in fact use the known $\alpha'$--corrections to the T-duality rules when doing the chain of T-duality--shift--T-duality. This strategy will automatically bring in the needed $\alpha'$--dependence into the deformed background, and will make sure that the deformed background is a solution to the two-loop equations. The interplay between T-duality and higher $\alpha'$--corrections was studied in various works~\cite{Panvel:1992he,Haagensen:1997er,Kaloper:1997ux,Jack:1999av,Parsons:1999ze,Edelstein:2019wzg}. In this paper we will use the $\alpha'$--corrections for the T-duality rules of~\cite{Kaloper:1997ux}, to obtain explicit $\alpha'$-corrections for YB deformed models. This strategy allows us to start from any background with isometries (it is not necessary to set the NSNS-flux to zero), and to keep the dependence on the deformation parameter exact.

Certain YB deformations, while they cannot be understood as simple TsT transformations, can still be obtained as a non-commuting sequence of TsT's \cite{Borsato:2016ose}. The non-commutativity is related to the fact that certain isometries needed to perform one TsT transformation may be broken by the application of another TsT. Therefore, in certain cases a sequence of TsT transformations can be implemented only in one precise order. Non-commuting sequences of TsT transformations are nice examples to study, because we can obtain explicit results by applying what is known about abelian T-duality and TsT, and at the same time be able to say something about NATD and more general YB deformations.

In the remaining part of the introduction, we will summarize the main results obtained when expanding the two-loop equations to second order in the deformation parameter.

\subsection{First \texorpdfstring{$\alpha'$}{alpha'}-correction to deformed backgrounds}
The (homogeneous) Yang-Baxter deformation of a bosonic string background $G,B,\Phi$ is given by \cite{Sakamoto:2017cpu,Araujo:2017jkb,Bakhmatov:2017joy,Borsato:2018idb}
\begin{equation}
\tilde G-\tilde B=(G-B)(1+\eta\Theta(G-B))^{-1}\,,\qquad\tilde\Phi=\Phi-\tfrac12\ln\det\left(1+\eta\Theta(G-B)\right)\,.
\label{eq:SWmap}
\end{equation}
Here $\eta$ is the deformation parameter and $\Theta$ is constructed by taking an anti-symmetric R-matrix solving the classical Yang-Baxter equation (CYBE), $R^{[I|L|}R^{J|M|}f_{LM}{}^{K]}=0$, on a subalgebra of the isometry algebra of the original background (with structure constants $f_{IJ}{}^K$) and contracting with the corresponding Killing vectors
\begin{equation}
\Theta^{ij}=k_I{}^iR^{IJ}k_J{}^j\equiv k^i\times k^j\,,\qquad\nabla_{(i}k_{Ij)}=0\,,
\label{eq:Theta}
\end{equation}
where we simplify the notation by introducing the anti-symmetric product '$\times$'. Assuming that $G,B,\Phi$ define a one-loop conformal bosonic string sigma model, the same is true of $\tilde G,\tilde B,\tilde\Phi$ if $R$ is unimodular, i.e. $R^{IJ}f_{IJ}{}^K=0$ \cite{Borsato:2016ose}.\footnote{The unimodularity condition is sufficient but not necessary in general. Relaxing it one finds at order $\eta$, assuming $B=0$, the necessary condition $dK=0$ where $K^n=\nabla_m\Theta^{mn}$. This is equivalent to $\nabla_mk_I^nf_{JK}{}^IR^{JK}=0$ which is in general weaker than the unimodularity condition $k_I^nf_{JK}{}^IR^{JK}=0$. The reason for this is that sometimes the anomalous terms generated by a non-unimodular $R$ can be removed by a field redefinition \cite{Wulff:2018aku} (see also \cite{Borsato:2018spz}). Here we will take $R$ to be unimodular for simplicity.} 

Here we want to ask what happens at two loops, i.e. the next order in $\alpha'$. We will work in an expansion in the deformation parameter up to order $\eta^2$. To simplify the calculations we will assume that the starting background has $B=0$ which gives the deformed background
\begin{equation}
\tilde G_{ij}=G_{ij}+\eta^2(\Theta^2)_{ij}+\mathcal O(\eta^4)\,,\qquad
\tilde B_{ij}=\eta\Theta_{ij}+\mathcal O(\eta^3)\,,\qquad
\tilde\Phi=\Phi-\tfrac14\eta^2\Theta_{ij}\Theta^{ij}+\mathcal O(\eta^4)\,.
\label{eq:deformed-bkg}
\end{equation}

We find that to this order in $\eta$ the first $\alpha'$--correction (i.e. two--loop correction) to the background is given by (in the scheme of Hull and Townsend \cite{Hull:1987yi})
{
\begin{align}
\delta\tilde G_{ij}=&\delta G_{ij}+2\eta^2(\delta G\Theta^2)_{(ij)}+\eta^2(\Theta\delta G\Theta)_{ij}
-2\eta^2\Theta_{k(i}R_{j)}{}^{klm}\Theta_{lm}+\eta^2\Theta^{mn}\nabla_i\nabla_j\Theta_{mn}
\,,
\nonumber\\
\delta\tilde B_{ij}=&2\eta(\delta G\Theta)_{[ij]}-\eta R_{ijkl}\Theta^{kl}\,,
\label{eq:corrections}
\\
\delta\tilde\Phi=&\delta\Phi-\tfrac12\eta^2(\delta G\Theta)_{mn}\Theta^{mn}
+\tfrac{1}{16}\eta^2\nabla^k\Theta^{mn}\nabla_k\Theta_{mn}-\tfrac{3}{8}\eta^2\nabla^k\Theta^{mn}\nabla_m\Theta_{nk}
+\tfrac14\eta^2\nabla_i\Phi\,\nabla^i(\Theta^{mn}\Theta_{mn})
\,.
\nonumber
\end{align}
}
Here $\delta G$, $\delta\Phi$ denote the $\alpha'$ corrections to the undeformed background with $B=\delta B=0$. Note that the terms involving $\delta G$ just come from correcting the undeformed metric in (\ref{eq:deformed-bkg}), while the terms involving the Riemann tensor in $\delta\tilde G$ and $\delta\tilde B$ are obtained simply by replacing $\Theta_{ij}\rightarrow\Theta_{ij}-\alpha'R_{ijkl}\Theta^{kl}$ in (\ref{eq:deformed-bkg}). 
{
The correction to the dilaton does not look nice in this scheme but by changing the scheme one can arrange it so that
\begin{equation}
e^{-2\tilde\Phi}\sqrt{\det\tilde G}=e^{-2\Phi}\sqrt{\det G}\,,
\label{eq:dilaton-inv}
\end{equation}
so that the correction to the dilaton just comes from the correction to the determinant of the metric. This is achieved by the scheme change\footnote{On-shell this is equivalent to turning on the $q$ parameter in the scheme of Hull and Townsend \cite{Hull:1987yi}.}
\begin{equation}
\Phi\rightarrow\Phi+\alpha'\left(-\tfrac12\nabla^2\Phi+(\nabla\Phi)^2-\tfrac{1}{16}H_{klm}H^{klm}\right)\,.
\label{eq:dilaton-scheme}
\end{equation}

With a little help from the corresponding expressions derived to all orders in $\eta$ for a particular background in (\ref{eq:BII-corr1}) and (\ref{eq:BII-corr2}) one can write a completion of (\ref{eq:corrections}) to all orders in the deformation. First of all it is natural to expect that one should correct the undeformed metric and take $\Theta_{ij}\rightarrow\Theta_{ij}-\alpha'R_{ijkl}\Theta^{kl}$ in the expressions in (\ref{eq:SWmap}). On top of this we need to extend the last term in the transformation of the metric and looking at the example in (\ref{eq:BII-corr1}) and (\ref{eq:BII-corr2}) suggests the following form for the corrections to all orders in $\eta$
\begin{align}
\tilde G_{ij}-\tilde B_{ij}
=&
\left[G(1+\eta[\Theta-\alpha'R\cdot\Theta])^{-1}\right]_{ij}
-\tfrac12\alpha'\partial_i\ln\det(1+\eta\Theta)\partial_j\ln\det(1+\eta\Theta)\,
\nonumber\\
&{}
+\tfrac12\alpha'\eta\left(\left[G(1+\eta\Theta)^{-1}\right]_{ik}\nabla^k\nabla_j\Theta^{mn}+\left[G(1-\eta\Theta)^{-1}\right]_{jk}\nabla^k\nabla_i\Theta^{mn}\right)\left[G(1+\eta\Theta)^{-1}\right]_{nm}
\label{eq:corrections-all-orders}
\end{align}
with the transformation of the dilaton read off from (\ref{eq:dilaton-inv}) (in the HT scheme after the shift (\ref{eq:dilaton-scheme})). Here indices are raised and lowered with the undeformed metric \emph{including} its $\alpha'$-corrections. We have also defined the contraction of $\Theta$ with the Riemann tensor $(R\cdot\Theta)_{ij}=R_{ijkl}\Theta^{kl}$. Note that this expression can be thought of as an $\alpha'$-corrected open-closed string map, such as appears for example in the work of Seiberg and Witten on non-commutative gauge theories \cite{Seiberg:1999vs}. While this result works for the rank 2 examples in section \ref{sec:Tdual} it unfortunately does not work in general.
}

\section{Two-loop conformal invariance conditions}
The conditions for two--loop conformal invariance of the bosonic string sigma model were worked out in \cite{Hull:1987pc,Metsaev:1987zx,Zanon:1987pp}. Following Hull and Townsend (HT) the conditions in their scheme are \cite{Hull:1987yi}\footnote{To go from their conventions to ours one sends $\Phi\rightarrow2\Phi$ and $H\rightarrow\frac12H$.}
\begin{equation}
\eqcf_{ij}^G=\eqcf_{0,ij}^G+\alpha'\eqcf_{1,ij}^G=0\,,\qquad
\eqcf_{ij}^B=\eqcf_{0,ij}^B+\alpha'\eqcf_{1,ij}^B=0\,,\qquad
\eqcf_{ij}^\Phi=\eqcf_{0,ij}^\Phi+\alpha'\eqcf_{1,ij}^\Phi=0\,,
\end{equation}
where the one-loop conditions are
\begin{equation}
\begin{aligned}
\eqcf_{0,ij}^G&=R_{ij}-\tfrac14H_{ikl}H_j{}^{kl}+2\nabla_i\nabla_j\Phi\,,\\
\eqcf_{0,ij}^B&=\nabla^kH_{ijk}-2\nabla^k\Phi H_{ijk}\,,\\
\eqcf_{0,ij}^\Phi&=2\nabla^2\Phi-4\nabla_i\Phi\nabla^i\Phi+\tfrac16 H_{ijk}H^{ijk}
\label{eq:eom-0}
\end{aligned}
\end{equation}
and the two-loop corrections are
\begin{align}
\eqcf_{1,ij}^G=&\,
\tfrac12R_{iklm}R_j{}^{klm}
+\tfrac14R_{iklj}H^{kmn}H^l{}_{mn}
+\tfrac14R_{klm(i}H_{j)}{}^{mn}H^{kl}{}_n
+\tfrac{1}{24}\nabla_iH_{klm}\nabla_jH^{klm}
\nonumber\\
&\,{}-\tfrac18\nabla^kH^{lm}{}_i\nabla_kH_{lmj}
+\tfrac{1}{16}H_{ikp}H_{jlq}H^{klm}H^{pq}{}_m
+\tfrac{1}{16}H_{ikp}H_{jl}{}^pH^{kmn}H^l{}_{mn}\,,
\label{eq:G-eom-1}
\\
&\nonumber\\
\eqcf_{1,ij}^B=&\,\nabla^kH^{lm}{}_{[i}R_{j]klm}
-\tfrac14\nabla_kH_{lij}H^{kmn}H^l{}_{mn}
+\tfrac12\nabla^kH^{lm}{}_{[i}H_{j]mn}H_{kl}{}^n\,,
\label{eq:B-eom-1}
\\
&\nonumber\\
\eqcf_{1,ij}^\Phi=&\,
-\tfrac14 R_{ijkl}R^{ijkl}+\tfrac{1}{12} (\nabla_iH_{jkl})(\nabla^iH^{jkl})+\tfrac18 H^{ij}{}_mH^{klm}R_{ijkl}+\tfrac14 R_{ij}(H^2)^{ij} \nonumber\\
&\,{}-\tfrac{5}{96}H_{ijk}H^i{}_{lm}H^{jl}{}_nH^{kmn}-\tfrac{3}{32}H^2_{ij}(H^2)^{ij},
\label{eq:Phi-eom-1}
\end{align}
where $H^2_{ij}=H_{ikl}H_j{}^{kl}$. Here we have set to zero the parameter $q$ of~\cite{Hull:1987yi}.

\section{Expansion in the deformation parameter}\label{sec:eta-exp}
In this section we expand the conditions for two-loop conformal invariance in powers of the deformation parameter $\eta$, and we find the explicit $\alpha'$ corrections for the background such that the conditions hold to the quadratic order in $\eta$. Here will not need to impose the equation for the dilaton. It is known that when the equations for $G$ and $B$ are satisfied the dilaton equation is satisfied up to a constant \cite{Hull:1987yi}. Since we assume the undeformed background to solve all the two-loop equations and since there is no way to introduce a constant at higher orders in $\eta$,\footnote{The parameter $\eta$ is always accompanied by $\Theta$ and it is not possible to construct a constant from a general $\Theta$.} the dilaton equation will not add anything.

\subsection{First order in the deformation parameter}
At order $\eta^1$ we see, by looking at (\ref{eq:deformed-bkg}), that the metric is not deformed while\footnote{We indicate the order in $\eta$ with a superscript in parenthesis. Since it is clear that this refers to the deformed background we drop the tilde.}
\begin{equation}
H^{(1)}_{ijk}=3\nabla_{[i}\Theta_{jk]}\,.
\end{equation}
Using this in (\ref{eq:B-eom-1}) we find
\begin{align}
\eqcf_{1,ij}^{B(1)}
=&\,
\nabla^kH^{(1)lm}{}_{[i}R_{j]klm}
=
\nabla^k(H^{(1)}_{lm[i}R_{j]k}{}^{lm})
+2H^{(1)}_{lm[i}\nabla^lR_{j]}{}^m
\nonumber\\
=&\,
\tfrac32\nabla^k\nabla_{[i}(R_{jk]lm}\Theta^{lm})
-\tfrac12\nabla^k[R_{ijlm}\nabla_k\Theta^{lm}]
+2\nabla_k(R_{[i}{}^{klm}\nabla_{|l|}\Theta_{j]m})
-2\nabla^k\Phi\,H^{(1)}_{lm[i}R_{j]k}{}^{lm}\,,
\end{align}
where we have used the lowest order equations (\ref{eq:eom-0}). Using the two derivative Killing identity (\ref{eq:2-der-id}) we have
\begin{align}
\nabla_k(R_i{}^{klm}\nabla_l\Theta_{jm})
=&
\nabla_kR_i{}^{klm}\nabla_l\Theta_{jm}
+R_i{}^{klm}\nabla_k\nabla_l\Theta_{jm}
\nonumber\\
=&
\nabla_kR_i{}^{klm}\nabla_l\Theta_{jm}
+2R_i{}^{klm}\nabla_k\nabla_{(l}\Theta_{j)m}
-R_i{}^{klm}\nabla_k\nabla_j\Theta_{lm}
\nonumber\\
=&
-\tfrac32\nabla_k(R_i{}^{klm}\nabla_j\Theta_{lm})
+2R_i{}^{klm}R_{jkln}\Theta_{mn}
-R_{im}{}^{kl}R_{jnkl}\Theta^{mn}
\nonumber\\
&{}
+R^{klmn}R_{klmi}\Theta_{jn}
+3R_{iklm}\nabla^k\Phi\,\nabla_j\Theta^{lm}
+2R_i{}^{klm}\nabla_k\Phi\,\nabla_l\Theta_{jm}\,.
\end{align}
Using this together with the identity (\ref{eq:id1}) we find
\begin{equation}
\eqcf_{1,ij}^{B(1)}
=
3\nabla^k\nabla_{[i}(R_{jk]lm}\Theta^{lm})
-6\nabla^k\Phi\,\nabla_{[i}(R_{jk]lm}\Theta^{lm})
+2R^{klmn}R_{klm[i}\Theta_{j]n}\,.
\end{equation}
Taking into account the $\alpha'$--corrections to the classical background, $\alpha'\delta G$ and $\alpha'\delta\Phi$, and the $B$-field at order $\eta^1$, $\alpha'(\delta\tilde B)^{(1)}$, we have
\begin{align}
\alpha'^{-1}\eqcf_{ij}^B
=&
3\nabla^k\nabla_{[i}(\delta\tilde B)^{(1)}_{jk]}
-6\nabla^k\Phi\,\nabla_{[i}(\delta\tilde B)^{(1)}_{jk]}
+3\nabla^k\nabla_{[i}(R_{jk]lm}\Theta^{lm})
-6\nabla^k\Phi\,\nabla_{[i}(R_{jk]lm}\Theta^{lm})
\nonumber\\
&{}
+3\delta(\nabla^k)\nabla_{[i}\Theta_{jk]}
-6\delta(\nabla^k\Phi)\,\nabla_{[i}\Theta_{jk]}
+2R^{klmn}R_{klm[i}\Theta_{j]n}\,.
\end{align}
In the case where the metric and dilaton do not receive corrections, $\delta G=\delta\Phi=0$, the terms in the second line vanish, and the terms in the first line also vanish provided we take
\begin{equation}
(\delta\tilde B)^{(1)}_{ij}=-R_{ijkl}\Theta^{kl}\,.
\end{equation}
In the general case the assumption that the corrected original background solves the two-loop equations implies that
\begin{align}
R^{klm}{}_nR_{klmi}
=&
-2\delta(R_{in}+2\nabla_i\nabla_n\Phi)
=
-\nabla^k\nabla_i\delta G_{kn}
-\nabla^k\nabla_n\delta G_{ki}
+G^{kl}\nabla_i\nabla_n\delta G_{kl}
\nonumber\\
&{}
+\nabla^2\delta G_{in}
+2\nabla^k\Phi(\nabla_i\delta G_{kn}+\nabla_n\delta G_{ki}-\nabla_k\delta G_{in})
-4\nabla_i\nabla_n\delta\Phi\,,
\label{eq:Riemann2}
\end{align}
where we used the expressions for the variation of the Ricci tensor and Christoffel symbols (\ref{eq:deltad2Phi}) and (\ref{eq:deltaRicci}).

Using this it is not hard to see, noting that $\delta\Phi$ must respect the isometries, that the $\delta\Phi$-terms cancel without any further correction to $B$. With a little bit more work one can show, using the fact that $\mathcal L_k\delta G_{ij}=0$, i.e. that the correction to the undeformed metric does not break any isometries, that all terms cancel if one takes
\begin{equation}
(\delta\tilde B)^{(1)}_{ij}=2(\delta G\Theta)_{[ij]}-R_{ijkl}\Theta^{kl}\,.
\label{eq:deltaB1}
\end{equation}
The first term is simply the correction induced by the correction to the undeformed metric, i.e. $\delta(B^{(1)})_{ij}=\delta\Theta_{ij}$, which comes from the fact that the indices on $\Theta_{ij}$ were lowered with the metric (note that the Killing vectors $k_I^m$, with an upper index, are not corrected by assumption). Thus we have proven that a two-loop Weyl invariant sigma-model remains two-loop Weyl invariant under a YB deformation, at least to first order in the deformation parameter. We now consider what happens at second order.

\subsection{Second order in the deformation parameter}
It is easy to see that at order $\eta^2$ the $B$-field equation, $\eqcf_{1,ij}^{B(2)}=0$, is trivially satisfied. For the metric equation we find
\begin{align}
\eqcf_{1,ij}^{G(2)}=&\,
R^{(2)}_{(i}{}^{klm}R_{j)klm}
-\tfrac12R_{(i}{}^{klm}R_{j)nlm}(\Theta^2)_k{}^n
-R_{(i}{}^{klm}R_{j)kl}{}^n(\Theta^2)_{mn}
+\tfrac14R_{kijl}H^{(1)kmn}H^{(1)l}{}_{mn}
\nonumber\\
&\,{}
+\tfrac14R_{klm(i}H^{(1)}_{j)}{}^{mn}H^{(1)kl}{}_n
+\tfrac{1}{24}\nabla_iH^{(1)}_{klm}\nabla_jH^{(1)klm}
-\tfrac18\nabla^kH^{(1)lm}{}_i\nabla_kH^{(1)}_{lmj}\,.
\label{eq:Phi2}
\end{align}
Note that we choose to define all tensors to have lower indices, e.g. $R_{ijkl}$, and then raise indices with the \emph{undeformed} metric $G_{ij}$.

The last two terms do not involve the Riemann tensor and the calculations can be simplified somewhat if we remove them by shifting the metric and dilaton. Under a shift of the metric we have
\begin{equation}
\delta(\nabla_i\nabla_j\Phi)
=
-\delta\Gamma_{ij}^k\nabla_k\Phi
=
-\tfrac12\nabla^k\Phi(\nabla_j\delta G_{ki}+\nabla_i\delta G_{kj}-\nabla_k\delta G_{ij})
\label{eq:deltad2Phi}
\end{equation}
and
\begin{equation}
\delta R_{ijkl}
=
\nabla_k(\delta\Gamma_{ilj}-\Gamma^m_{lj}\delta G_{im})
+\tfrac12R^m{}_{jkl}\delta G_{im}
-(k\leftrightarrow l)\,,
\end{equation}
so that in particular
\begin{align}
R^{(2)}_{ijkl}=&
\nabla_k(\Gamma^{(2)}_{[ij]l}+\Gamma^m_{l[i}(\Theta^2)_{j]m})
-\tfrac12(\Theta^2)^m{}_{[i}R_{j]mkl}
-(k\leftrightarrow l)
\nonumber\\
=&
-\nabla_k\nabla_{[i}(\Theta^2)_{j]l}
+\nabla_l\nabla_{[i}(\Theta^2)_{j]k}
-(\Theta^2)^m{}_{[i}R_{j]mkl}\,.
\end{align}
The variation of the Ricci tensor becomes (symmetrization in $ij$ understood)
\begin{align}
\delta R_{ij}
=&
\delta G^{kl}R_{ikjl}
+G^{kl}\delta R_{ikjl}
\nonumber\\
=&
\delta G_{kl}R^k{}_{ij}{}^l
+R^k{}_j\delta G_{ik}
+\nabla_j[G^{kl}\delta\Gamma_{ikl}-G^{kl}\Gamma^m_{kl}\delta G_{im}]
-\nabla^k[\delta\Gamma_{ijk}-\Gamma^l_{jk}\delta G_{il}]
\nonumber\\
=&
\nabla^k\nabla_i\delta G_{kj}
-\tfrac12G^{kl}\nabla_i\nabla_j\delta G_{kl}
-\tfrac12\nabla^2\delta G_{ij}\,.
\label{eq:deltaRicci}
\end{align}
From this expression we see that the last two terms in (\ref{eq:Phi2}) can be canceled by shifting the metric and dilaton as
\begin{equation}
G_{ij}\rightarrow G_{ij}-\tfrac18\alpha'H_{ikl}H_j{}^{kl}\,,\qquad\Phi\rightarrow\Phi-\tfrac{1}{32}\alpha'H_{klm}H^{klm}\,.
\label{eq:scheme-shift}
\end{equation}
The two-loop contribution then becomes (symmetrization in $ij$ understood)
\begin{align}
\eqcf_{1,ij}'^{G(2)}=&\,
R^{(2)}_i{}^{klm}R_{jklm}
-\tfrac12R_i{}^{klm}R_{jnlm}(\Theta^2)_k{}^n
-R_i{}^{klm}R_{jkl}{}^n(\Theta^2)_{mn}
+\tfrac18R_{kijl}H^{(1)kmn}H^{(1)l}{}_{mn}
\nonumber\\
&\,{}
+\tfrac12R_{klmi}H^{(1)}_j{}^{mn}H^{(1)kl}{}_n
-\tfrac18R^{klmn}H^{(1)}_{ikl}H^{(1)}_{jmn}
-\tfrac{1}{24}H^{(1)klm}\nabla_i\nabla_jH^{(1)}_{klm}\,.
\end{align}
Here we have used the Bianchi identity for $H$ and the lowest order equations of motion, which in particular imply
\begin{equation}
\nabla^2H_{klm}=3\nabla^n\nabla_{[k}H_{lm]n}=-3R_{np[kl}H_{m]}{}^{np}+6\nabla^n\Phi\,\nabla_{[k}H_{lm]n}\,.
\end{equation}
Note that terms with two derivatives of $H^{(1)}$ indeed give something involving the Riemann tensor since they involve three derivatives acting on a product of two Killing vectors giving at least two derivatives on one Killing vector.

Expressing all terms in terms of the basis defined in appendix \ref{app:relations} we have (symmetrization in $ij$ understood)
\begin{align}
R^{(2)}_i{}^{klm}R_{jklm}
=&
-\nabla\cdot(f_{12}+f_{20})
-\nabla(2\hat f_5-\hat f_6)
+2g_{32}
+g_{34}
-g_{35}
+h_7
-\tfrac12h_8
\nonumber\\
&{}
+\tfrac12h_{10}
+2m_7+2m_9
\\
R_{klmi}H^{(1)}_j{}^{mn}H^{(1)kl}{}_n
=&
-g_3
+2g_4
-2g_6
+g_8
-2g_{14}
+g_{15}
\\
R^{klmn}H^{(1)}_{ikl}H^{(1)}_{jmn}
=&
4g_{16}
+4g_{17}
+g_{19}\,,
\\
H^{(1)klm}\nabla_i\nabla_jH^{(1)}_{klm}
=&
3g_3
-6g_4
-6g_6
+3g_8
-18g_{14}
+9g_{15}
+6g_{28}
-6g_{29}
-3g_{31}
+12g_{32}
\nonumber\\
&{}
-12g_{33}
+12g_{34}\,.
\end{align}

While the order $\eta$ $\alpha'$--correction to $\tilde B$ in (\ref{eq:deltaB1}) contributes the terms (for the moment we assume that the undeformed background is not corrected) (symmetrization in $ij$ understood)
\begin{equation}
-\tfrac12(\delta\tilde H)^{(1)}_{ikl}H^{(1)}_j{}^{kl}
=
\tfrac32\nabla_{[i}(R_{kl]mn}\Theta^{mn})H^{(1)}_j{}^{kl}
=
g_3
-g_8
-g_{15}
+g_{16}
+\tfrac12g_{19}\,.
\end{equation}

For the two-loop correction we therefore get $\frac18$ times
\begin{align}
&{}
-8\nabla\cdot(f_{12}+f_{20})%
-8\nabla(2\hat f_5-\hat f_6)%
+3g_3%
+10g_4%
-6g_6%
-5g_8%
-2g_{14}%
-7g_{15}%
+4g_{16}%
-4g_{17}%
\nonumber\\
&{}
+3g_{19}%
+4g_{30}
+2g_{31}
+12g_{32}
+4g_{33}
+4g_{34}
-8g_{35}
-8h_8%
+4h_{10}
+16m_7
+16m_9
\end{align}
To this we have to add the terms arising from the $\alpha'$--corrections to $\tilde G$ and $\tilde\Phi$. We will ignore the corrections to the undeformed background until the end of the section.

Consider the following possible $\alpha'$--corrections to the metric at order $\eta^2$ (symmetrization in $ij$ understood)
\begin{align}
\delta_1\tilde G_{ij}=&\,\nabla_i\Theta_{mn}\nabla_j\Theta^{mn}\,,\\
\delta_2\tilde G_{ij}=&\,k_i\times\nabla_mk_n\,k_j\times\nabla^mk^n\,,\\
\delta_3\tilde G_{ij}=&\,\nabla_i\Theta^{mn}\nabla_m\Theta_{nj}\,,\\
\delta_4\tilde G_{ij}=&\,R_i{}^{klm}\Theta_{jk}\Theta_{lm}\,.
\end{align}
Note that we could write also the second one in terms of $\Theta$ as
\begin{equation}
\delta_2\tilde G_{ij}=\tfrac12\nabla_m\Theta_{in}\nabla^m\Theta_j{}^n
-\tfrac12\nabla^n\Theta_{im}\nabla^m\Theta_{jn}
-\nabla_i\Theta^{mn}\nabla_m\Theta_{nj}
+\tfrac14\nabla_i\Theta^{mn}\nabla_j\Theta_{mn}\,,
\end{equation}
but the above expression is more convenient for the following calculation. Using (\ref{eq:deltaRicci}) and (\ref{eq:deltad2Phi}) these variations give rise to the terms
\begin{align}
\delta_1\tilde G:\quad&
-\nabla\cdot(2f_3+f_{28})
-\nabla(\hat f_1+2\hat f_6)
+g_{31}
-4m_5
-4m_6
+2m_{20}
\nonumber\\
&\nonumber\\
\delta_2\tilde G:\quad&
\tfrac12\nabla\cdot(f_1+2f_7-f_{14}-2f_{17}+f_{22}+2f_{23})%
+\nabla(-\hat f_1+2\hat f_2+2\hat f_3-\hat f_4+2\hat f_5)%
+g_{28}
\nonumber\\
&{}
-g_{29}
-2g_{30}
+\tfrac12g_{31}
-2m_{12}
+m_{13}
+\tfrac38\nabla_i\nabla_j(2\nabla^k\Theta^{mn}\nabla_m\Theta_{nk}-3\nabla^k\Theta^{mn}\nabla_k\Theta_{mn})
\nonumber\\
&\nonumber\\
\delta_3\tilde G:\quad&
-\tfrac12\nabla\cdot(f_1+f_3+f_{10}-f_{11}+f_{22}+f_{28}+f_{30}-f_{31})%
\nonumber\\
&{}
+\tfrac14\nabla(
\hat f_1
-2\hat f_2
-2\hat f_3
+\hat f_4
-2\hat f_5
+2\hat f_7
-4\hat f_8
)
+g_{30}
-m_5
-m_6
+m_7
-m_8
-m_{10}
\nonumber\\
&{}
+m_{11}
-m_{13}
+m_{20}
+m_{22}
-m_{23}
\nonumber\\
&\nonumber\\
\delta_4\tilde G:\quad&
\tfrac12\nabla\cdot(f_9+f_{14}-f_{26})
-\tfrac14\nabla
(
3\hat f_1
+2\hat f_2
+2\hat f_3
-3\hat f_4
-2\hat f_5
+4\hat f_6
)
+h_9
-m_1
+m_2
\nonumber\\
&{}
-m_3
-m_{15}\,,
\nonumber
\end{align}
where we used the identity (\ref{eq:id04}) in calculating the last variation.

Taking the following correction to the metric and dilaton
\begin{equation}
(\delta\tilde G)^{(2)}_{ij}=\tfrac14(-3\delta_1+2\delta_2+2\delta_3+6\delta_4)\tilde G_{ij}\,,\qquad(\delta\tilde\Phi)^{(2)}=-\tfrac{3}{32}(2\nabla^k\Theta^{mn}\nabla_m\Theta_{nk}-3\nabla^k\Theta^{mn}\nabla_k\Theta_{mn})
\end{equation}
and using appendix \ref{app:relations} we are left with $\frac18$ times the following order $\alpha'$ terms
\begin{align}
&{}
12g_1
+8g_2
+g_3
-6g_4
+4g_5
-6g_6
-12g_7
+3g_8
+12g_{10}
-9g_{12}
\nonumber\\
&{}
+24g_{13}
+12g_{14}
-6g_{15}
+6g_{16}
-6g_{19}
-6g_{20}
+12g_{21}
+8g_{22}
-2g_{23}
-12g_{24}
+16g_{25}
\nonumber\\
&{}
+6h_1
+8h_2
-16h_3
-4h_5
+16h_6
-4h_8
+12h_9
+8h_{10}
-4h_{11}
+4\nabla\hat f_7
\end{align}
Next we use the Yang-Baxter equation which, in terms of $\Theta$, reads
\begin{equation}
\Theta^{k[l}\nabla_k\Theta^{mn]}=0\,.
\end{equation}
Hitting this with $R_{ipmn}\nabla^p$ we get the identity
\begin{equation}
0=
R_{ilmn}\nabla^l(\Theta_{kj}\nabla^k\Theta^{mn})
+2R_i{}^{lmn}\nabla_l(\Theta_{km}\nabla^k\Theta_{nj})
=
\nabla\cdot(f_{19}-2f_{11})\,.
\end{equation}
Adding $-4$ times the RHS to our expression we are left with $\frac18$ times
\begin{align}
&{}
12g_1
+8g_2
-3g_3
-6g_4
+12g_5
-6g_6
-12g_7
+3g_8
+12g_{10}
-9g_{12}
\nonumber\\
&{}
+24g_{13}
+12g_{14}
-6g_{15}
+6g_{16}
-6g_{19}
-6g_{20}
+8g_{21}
+8g_{22}
-6g_{23}
-12g_{24}
+24g_{25}
\nonumber\\
&{}
+6h_1
+8h_2
-16h_3
+24h_6
+12h_9
+8h_{10}
-4h_{11}
-8(m_4-2m_{10}+2m_{11})
+4\nabla\hat f_7\,,
\end{align}
where the $m$-terms vanish by the Yang-Baxter equation. Using the identities (\ref{eq:id03})--(\ref{eq:id01}), (\ref{eq:f14id}) and (\ref{eq:h5id}) this reduces to (symmetrization in $ij$ understood)
\begin{equation}
h_{10}
-\tfrac12h_{11}
+\tfrac12\nabla\hat f_7
=
R_{klmi}R^{klmn}(\Theta^2)_{nj}
-\tfrac12R_{klm}{}^nR^{klmp}\Theta_{in}\Theta_{jp}
+\tfrac14\nabla_i\nabla_j(\nabla^l\Theta^{mn}\nabla_l\Theta_{mn})
\,.
\end{equation}
The first two terms vanish if the original background does not suffer $\alpha'$--corrections, while the last term can be canceled by shifting the dilaton.

To summarize we have found that with the following correction to the metric and dilaton in the HT scheme at order $\eta^2$, taking into account also (\ref{eq:scheme-shift}), (symmetrization in $ij$ understood)
\begin{align}
(\delta\tilde G)^{(2)}_{ij}=&-\tfrac34\nabla_i\Theta^{mn}\nabla_j\Theta_{mn}-\tfrac12\nabla_m\Theta_{ni}\nabla_j\Theta^{mn}-\tfrac32R_i{}^{klm}\Theta_{lm}\Theta_{kj}\,,\\
(\delta\tilde\Phi)^{(2)}=&\tfrac{1}{16}\nabla^k\Theta^{mn}\nabla_k\Theta_{mn}-\tfrac{3}{8}\nabla^k\Theta^{mn}\nabla_m\Theta_{nk}\,,
\end{align}
the deformed model is Weyl invariant at two loops provided the undeformed model is. The shift in the metric does not look particularly natural but it can be brought to a nicer form by noting that (symmetrization in $ij$ understood)
\begin{equation}
\nabla_m\Theta_{ni}\nabla_j\Theta^{mn}=
\nabla_mk_n\times k_i\nabla_j\Theta^{mn}
+\tfrac12\nabla_i\Theta^{mn}\nabla_j\Theta_{mn}
=
\nabla_iv_j
+R_i{}^{klm}\Theta_{kj}\Theta_{lm}
+\tfrac12\nabla_i\Theta^{mn}\nabla_j\Theta_{mn}\,,
\label{eq:diffeo}
\end{equation}
where $v_j=\nabla_mk_n\times k_j\Theta^{mn}$. {The first term represents a diffeomorphism, so it can be dropped (note that the dilaton does not transform, $v^i\nabla_i\Phi=0$, since it is isometric). It will be convenient to perform a further diffeomorphism generated by $v^i=\frac12\Theta^{mn}\nabla^i\Theta_{mn}$ after which we have (symmetrization in $ij$ understood)
\begin{align}
(\delta\tilde G)^{(2)}_{ij}=&-2R_i{}^{klm}\Theta_{lm}\Theta_{kj}+\Theta^{mn}\nabla_i\nabla_j\Theta_{mn}
\,,\\
(\delta\tilde\Phi)^{(2)}=&\tfrac{1}{16}\nabla^k\Theta^{mn}\nabla_k\Theta_{mn}-\tfrac{3}{8}\nabla^k\Theta^{mn}\nabla_m\Theta_{nk}
+\tfrac14\nabla_i\Phi\,\nabla^i(\Theta^{mn}\Theta_{mn})
\,.
\end{align}
We will now consider what happens when the undeformed background receives $\alpha'$--corrections.
}

Taking into account the lowest order correction to the metric and dilaton as well as the first order correction to $\tilde B$ (\ref{eq:deltaB1}) we have (symmetrization in $ij$ understood)
\begin{align}
&{}
\delta(R^{(2)}_{ij}-\tfrac14H^{(1)}_{ikl}H^{(1)}_j{}^{kl}+2[\nabla_i\nabla_j\Phi]^{(2)})
+R_{klmi}R^{klmn}(\Theta^2)_{nj}
-\tfrac12R_{klm}{}^nR^{klmp}\Theta_{in}\Theta_{jp}\,.
\end{align}
Using (\ref{eq:Riemann2}) and the variations in (\ref{eq:deltaRicci}) and (\ref{eq:deltad2Phi}) this becomes, after a tedious calculation,
\begin{align}
&{}
-3\nabla^k\delta G^n{}_i\nabla^l\Theta_{[nk}\Theta_{j]l}%
-\delta G_{in}k^k\times[k^l,\nabla_lk_j]\times\nabla_kk^n%
+2\delta G_{kn}k^k\times[k^l,\nabla_lk_j]\times\nabla_ik^n%
\nonumber\\
&{}
+\delta G_{kn}\nabla_ik^k\times[k^l,\nabla_lk^n]\times k_j%
-\delta G_{kn}\nabla^n(k_i\times[k^l,\nabla_lk_j]\times k^k)%
\nonumber\\
&{}
-2\nabla_k\Phi\,\delta G_{in}k^n\times[k^l,\nabla_lk^k]\times k_j
+2\nabla^k\Phi\,\delta G_{kn}k_i\times[k^l,\nabla_lk^n]\times k_j\,.
%
\end{align}
The first term vanishes by the Yang-Baxter equation. Using the fact that $k_I^l\nabla_lk_J^n-k_J^l\nabla_lk_I^n=f_{IJ}{}^Kk_K^n$ and the YB equation (i.e. $R^{IJ}R^{KL}f_{JK}{}^M$ antisymmetrized in $ILM$ vanishes) this further reduces to
\begin{align}
-\tfrac12R^{IJ}R^{KL}f_{JK}{}^Mf_{IL}{}^N\delta G_{in}k_{Mj}k_N^n
=&
R^{MJ}R^{KI}f_{JK}{}^Lf_{IL}{}^N\delta G_{in}k_{Mj}k_N^n
\nonumber\\
=&
-\tfrac12R^{MJ}R^{KI}f_{KI}{}^Lf_{JL}{}^N\delta G_{in}k_{Mj}k_N^n
=0\,,
\end{align}
where we have used first the YB equation, then the Jacobi identity and finally the unimodularity condition $R^{KI}f_{KI}{}^L=0$.

This shows that the only additional corrections that arise are the ones coming from correcting the undeformed metric in $\tilde G^{(2)}$ and $\Phi^{(2)}$ so that
{
\begin{align}
(\delta\tilde G)^{(2)}_{ij}=&2(\delta G\Theta^2)_{(ij)}+(\Theta\delta G\Theta)_{ij}-2\Theta_{k(i}R_{j)}{}^{klm}\Theta_{lm}
+\Theta^{mn}\nabla_i\nabla_j\Theta_{mn}
\,,\\
(\delta\tilde\Phi)^{(2)}=&
-\tfrac12(\delta G\Theta)_{mn}\Theta^{mn}
+\tfrac{1}{16}\nabla^k\Theta^{mn}\nabla_k\Theta_{mn}-\tfrac{3}{8}\nabla^k\Theta^{mn}\nabla_m\Theta_{nk}
+\tfrac14\nabla_i\Phi\,\nabla^i(\Theta^{mn}\Theta_{mn})
\,.
\end{align}
}
This completes the proof that, at least to second order in the deformation and when $B=0$, unimodular YB deformations preserve conformality at two loops.

\section{\texorpdfstring{$\alpha'$}{alpha'}-corrections from T-duality rules at two loops}\label{sec:Tdual}
Homogeneous Yang-Baxter deformations are closely related to non-abelian T-duality~\cite{Hoare:2016wsk,Borsato:2016pas} and it can be shown that the non-abelian T-dual model is in fact recovered in the maximally deformed limit $\eta\to \infty$~\cite{Borsato:2016pas}, see also~\cite{Borsato:2017qsx,Borsato:2018idb}.
The simplest class of Yang-Baxter deformations --- the ``abelian'' one --- is related to just abelian T-duality, and is equivalent to doing TsT transformations~\cite{Frolov:2005dj,Alday:2005ww}. In general, a Yang-Baxter deformation generated by $\Theta=k_1\wedge k_2$ where $k_1=\partial_{x^1}$ and $k_2=\partial_{x^2}$ are commuting Killing vectors, is equivalent to doing first a T-duality $x^1\to \tilde x^1$, then a shift $x^2\to x^2+\eta \tilde x^1$, and then a T-duality back $\tilde x^1\to x^1$. Some ``non-abelian'' deformations are non-commuting sequences of TsT's~\cite{Borsato:2016ose,vanTongeren:2016eeb}. The non-abelian nature is related to the fact that the order in which the TsT transformations are performed is important, as certain T-dualities would break the isometries that are needed to perform the other T-dualities in the sequence. In this section we want to exploit the relation to TsT transformations and combine it with the knowledge of the first $\alpha'$--corrections of the T-duality rules, to obtain two-loop corrections for all Yang-Baxter deformations that are obtainable by TsT transformations, or more generically by a non-commuting sequence of them. This strategy allows us to obtain backgrounds at two loops that are exact in the deformation parameter $\eta$. Moreover, these tools can be applied to any starting background with isometries, and it is not needed to restrict to $B=0$ as we assume in most of this paper.

Because at each step all that we are doing is (abelian) T-duality and coordinate transformations, we are bound to preserve conformal invariance on the worldsheet to the very end, and we can check explicitly that the solutions we generate do solve the two-loop equations. This argument can be repeated also to higher orders in the $\alpha'$ expansion, and it is enough to conclude that all Yang-Baxter deformations that are obtainable by a generically non-commuting sequence of TsT transformations, do not break the conformality of the original model to all orders in $\alpha'$.

At leading order in $\alpha'$ the T-duality rules are given by the Buscher rules~\cite{Buscher:1987sk}. At higher loops these rules get corrected in $\alpha'$. We will use the $\alpha'$--corrections to the T-duality rules derived by Kaloper and Meissner in~\cite{Kaloper:1997ux}. The rules were obtained by carefully analysing the two-loop effective action of the bosonic string, and identifying the terms that are symmetric or anti-symmetric under the Buscher rules. The $\alpha'$--corrections of the T-duality rules were then fixed by requiring that they give a symmetry of the full two-loop effective action, compensating for the antisymmetry of those terms.\footnote{In~\cite{Kaloper:1997ux} the authors claim that their results can be applied also to the heterotic string, but the action they start with is missing the Chern-Simons terms that are expected there. See~\cite{Edelstein:2019wzg} for $\alpha'$--corrected T-duality rules that encompass both the bosonic and the heterotic string.}

Already at leading order in $\alpha'$, the T-duality rules are more easily presented in terms of fields of a dimensional reduction, where we reduce along the direction that we want to T-dualize. We follow~\cite{Kaloper:1997ux} and we rewrite the metric, Kalb-Ramond field and dilaton of the $D$-dimensional spacetime in terms of the following $(D-1)$-dimensional fields 
\be\label{eq:dim-red}
\begin{aligned}
ds^2 &= G_{ij}dx^idx^j=g_{\mu\nu}dx^\mu dx^\nu +e^{2\sigma}(d\xt+V)^2\,,\\
B&= \tfrac{1}{2}B_{ij}dx^i\wedge dx^j =\tfrac{1}{2}b_{\mu\nu}dx^\mu\wedge dx^\nu+\tfrac{1}{2}W\wedge V + W \wedge d\xt\,,\\
\Phi&=\phi+\tfrac{1}{2}\sigma\,.
\end{aligned}
\ee 
Here we are assuming that we have brought the solution in a form such that the isometry we want to dualize is simply implemented by a shift of a coordinate, that we denote by $\xt$. We use Greek indices for the $(D-1)$-dimensional spacetime.\footnote{The discussion of the $\alpha'$--corrected T-duality rules and their derivation simplifies if written in terms of tangent-space indices, but we will not do so here.} We have introduced a $(D-1)$-dimensional metric $g_{\mu\nu}$, and antisymmetric $b_{\mu\nu}$, vectors $V_\mu$ and $W_\mu$, and scalars $\phi$ and $\sigma$. Above we also used form notation $V=V_\mu dx^\mu ,W=W_\mu dx^\mu $. In components, the relations to identify the fields of the dimensional reduction are
\be
\begin{aligned}
& \sigma=\tfrac{1}{2}\log G_{\xt\xt }\,,\qquad
&& V_\mu =\frac{G_{\mu \xt }}{G_{\xt \xt }}\,,\qquad
&&&g_{\mu \nu }=G_{\mu \nu }-\frac{G_{\mu \xt }G_{\nu \xt }}{G_{\xt \xt }}\,,\\
&\phi=\Phi-\tfrac{1}{4}\log G_{\xt \xt }\, ,\qquad
&&W_\mu = B_{\mu \xt }\,,
&&&b_{\mu \nu } = B_{\mu \nu }+\frac{G_{\xt [\mu }B_{\nu ]\xt }}{G_{\xt \xt }}\,.
\end{aligned}
\ee
It is also useful to notice that $G^{\mu\nu}=g^{\mu\nu}\,,\quad G^{\mu\xt}=-V^\mu\,,\quad G^{\xt\xt} = e^{-2\sigma}+V^2\,.$
The combination 
\be
h_{\mu\nu\rho}=3(\partial_{[\mu}b_{\nu\rho]}-\tfrac12 W_{[\mu\nu}V_{\rho]}-\tfrac12 V_{[\mu\nu}W_{\rho]})=H_{\mu\nu\rho}-3 W_{[\mu\nu}V_{\rho]}\,,
\ee
is gauge invariant. In terms of these new fields the Buscher rules are simply
\be
\sigma\to -\sigma,\qquad\qquad\qquad V\leftrightarrow W\,.
\ee
All other fields remain unchanged under T-duality at leading order in $\alpha'$.

In~\cite{Kaloper:1997ux} Kaloper and Meissner derived the corrections to the T-duality rules in a particular scheme introduced by Meissner in~\cite{Meissner:1996sa}. We will call it the Kaloper-Meissner (KM) scheme. In order to apply the T-duality rules of KM to our case, we will therefore first need to implement the field redefinitions to go from the scheme of HT to that of KM. We can do so by combining the formulas given in~\cite{Hull:1987yi} (see their equations (61) and (64)) relating the HT scheme to the Metsaev-Tseytlin (MT) scheme of~\cite{Metsaev:1987zx},   and those given in~\cite{Meissner:1996sa} (see his equations (3.7), (4.1) and (4.7)) to go from MT to KM.\footnote{The field redefinitions given in~\cite{Meissner:1996sa} relate the KM and the MT schemes only on-shell, but this is enough for our purposes, since we just want to make sure that we can generate solutions of the two-loop equations.} The field redefinitions that we will use are\footnote{These are the redefinitions needed when we set the parameter $q$ of~\cite{Hull:1987yi} to zero. Different values of $q$ would affect the coefficient of $H^2$ that appears in the redefinition of the dilaton. Importantly, the coefficient in front of $H^2_{ij}$ that appears in the redefinition of the metric has the opposite sign compared to what one would expect from formulas in~\cite{Hull:1987yi} or ~\cite{Meissner:1996sa}. We have checked in various examples,  some not included in this paper, that we must have the sign that we use here, as this is fixed by requiring that we want to have a solution of the two-loop equations  after doing T-duality in the KM scheme and going back to the HT scheme.}
\be\label{eq:HT-KM}
\begin{aligned}
&G^{(\text{HT})}_{ij}=G^{(\text{KM})}_{ij}+\alpha'( R_{ij}-\tfrac{1}{2} H^2_{ij})\,,\\
&B^{(\text{HT})}_{ij}=B^{(\text{KM})}_{ij}+\alpha'(- H_{ijk}\nabla^k\Phi)\,,\\
&\Phi^{(\text{HT})}=\Phi^{(\text{KM})}+\alpha'(-\tfrac{3}{32}H^2+\tfrac{1}{8}R-\tfrac{1}{2} (\nabla\Phi)^2)\,.
\end{aligned}
\ee
Once we are in the scheme of KM we can use their $\alpha'$--corrected T-duality rules~\cite{Kaloper:1997ux}
\begin{eqnarray}
\sigma &\rightarrow&  -\sigma + \alpha' \left[
(\nabla \sigma)^2 + \tfrac{1}{8} (e^{2\sigma} Z + e^{-2\sigma} T)\right] 
 \nonumber\\
V_{\mu } &\rightarrow&  W_{\mu } + \alpha' \left[W_{\mu \nu } \nabla^{\nu } \sigma +\tfrac{1}{4}  h_{\mu \nu \rho }V^{\nu \rho } e^{2\sigma} \right] \label{eq:T-KM} \\
W_{\mu } &\rightarrow&  V_{\mu } + \alpha'\left[ V_{\mu \nu } \nabla^{\nu } \sigma
- \tfrac{1}{4}  h_{\mu \nu \rho } W^{\nu \rho } e^{-2\sigma}\right]\nonumber \\
b_{\mu \nu } &\rightarrow&  b_{\mu \nu }
+ \alpha' [V_{\rho [\mu }W^\rho {}_{\nu ]}+(W_{[\mu \rho }\nabla^\rho \sigma+\tfrac14 e^{2\sigma}h_{[\mu \rho \lambda }V^{\rho \lambda })V_{\nu ]}+(V_{[\mu \rho }\nabla^\rho \sigma-\tfrac14 e^{-2\sigma} h_{[\mu \rho \lambda }W^{\rho \lambda })W_{\nu ]}]\nonumber
\end{eqnarray}
Indices are always raised/lowered using the $(D-1)$-dimensional metric $g_{\mu\nu}$, and the transformations are written using also the following definitions
\be
\begin{aligned}
&V_{\mu\nu}=\partial_\mu V_\nu -\partial_\nu V_\mu\,,\qquad\qquad
&&Z_{\mu\nu}=V_{\mu \rho}V_\nu ^{\ \rho}\,,\qquad\qquad
&&&Z=Z_\mu ^{\ \mu }\,,\\
&W_{\mu\nu}=\partial_\mu W_\nu -\partial_\nu W_\mu\,,
&&T_{\mu\nu}=W_{\mu \rho}W_\nu ^{\ \rho}\,,
&&&T=T_\mu ^{\ \mu }\,.
\end{aligned}
\ee
In general, at higher loops, not only $\sigma, V$ and $W$ will change under T-duality. In fact, at two loops in the scheme of KM also $b_{\mu\nu}$ gets modified.\footnote{In~\cite{Kaloper:1997ux} the rules were given in terms of transformations of $h_{\mu\nu\rho}$. Here we preferred to rewrite them as a transformation of $b_{\mu\nu}$. {Importantly, the $\alpha'$-corrections to the T-duality rules of $b_{\mu\nu}$ (or equivalently $h_{\mu\nu\rho}$) differ by an overall sign compared to those given in~\cite{Kaloper:1997ux}, and our formula corrects the one given there. We thank A. Vilar L\'opez for discussions on this point.  A future paper will contain also more details on this~\cite{nextpaper}.}}
It is important to remark that already before doing T-duality the fields  will in general have an explicit $\alpha'$-dependence. In particular, $\sigma, V$ and $W$ that transform according to~\eqref{eq:T-KM} may in general depend on $\alpha'$, and this must be taken into account already when implementing the leading order T-duality rules (the Buscher rules).

One could in principle combine the T-duality rules of KM in~\eqref{eq:T-KM} with the field redefinitions in~\eqref{eq:HT-KM}, to obtain the $\alpha'$--corrections of the T-duality rules in the scheme of HT. We will not do so here, as the scheme of KM appears to be the minimal scheme for what concerns the complexity of the corrections to the T-duality rules. In other schemes, all other fields of the dimensional reduction will in general receive $\alpha'$--corrections.
Therefore, to obtain Yang-Baxter deformations in the scheme of HT we will follow this strategy:
\begin{enumerate}
\item Start from a solution of the two-loop equations in the HT scheme. In general that implies finding $\alpha'$--corrections for this initial solution.
\item Go to the scheme of KM  using~\eqref{eq:HT-KM}.
\item Do TsT or sequences of TsT transformations, using the $\alpha'$--corrected T-duality rules in~\eqref{eq:T-KM}.
\item Go back to the scheme of HT using~\eqref{eq:HT-KM}.
\end{enumerate}
We have worked out examples to test this method and obtain explicit results for $\alpha'$--corrections of Yang-Baxter deformed models. This also allows us to relate to the results of section~\ref{sec:eta-exp} that are perturbative in $\eta$. We will provide an example in the next section.

\section{Examples}
In this section we consider two particularly simple examples.

\subsection{Solvable pp-wave}
We start with the pp-wave background considered in \cite{Papadopoulos:2002bg}
\begin{equation}
ds^2=2dx^+dx^--\frac{k}{(x^+)^2}x_m^2(dx^+)^2+dx_m^2\,,\qquad
\Phi=mx^++\frac{d}{2}k\ln x^+\,,
%
%
%
%
\end{equation}
where $0<k<\frac14$ is a constant, $m$ is another constant and $d$ is the number of transverse dimensions. This background is known not to receive $\alpha'$--corrections. This follows from the fact that the only non-zero component of the Riemann tensor is $R_{+m+n}=\delta_{mn}k(x^+)^{-2}$.

Consider the following four Killing vectors
\begin{align}
k_1=&(x^+)^\nu\partial_1-\nu(x^+)^{\nu-1}x_1\partial_-\,, & k_3=&(2\nu-1)\partial_-\,,
\nonumber\\
k_2=&(x^+)^{1-\nu}\partial_1-(1-\nu)(x^+)^{-\nu}x_1\partial_-\,, & k_4=&(x^+)^\nu\partial_2-\nu(x^+)^{\nu-1}x_2\partial_-\,,
\end{align}
where we have defined the parameter
\begin{equation}
\nu=\frac{1+\sqrt{1-2k}}{2}\,.
%
\end{equation}
They form a Heisenberg algebra of isometries with the only non-trivial Lie bracket $[k_1,k_2]=k_3$. From the discussion of R-matrices in \cite{Borsato:2016ose} we see that we can consider the non-abelian rank 4 deformation
\begin{equation}
\Theta=k_1\wedge k_4+sk_2\wedge k_3\,,
\end{equation}
where we introduced the parameter $s$ to keep track of the contribution from the second term. We will show below that in this case this deformation is equivalent to the abelian one obtained by setting $s=0$. First we construct the matrix
\begin{equation}
\Theta^{ij}=\left(
\begin{array}{cccc}
0 & 0 & 0 & 0\\
0 & 0 & a & b\\
0 & -a &0 & c\\
0 & -b &-c& 0
\end{array}
\right)\,,
\end{equation}
where
\begin{equation}
a=\nu(x^+)^{2\nu-1}x_2-s(2\nu-1)(x^+)^{1-\nu}\,,\qquad
b=-\nu(x^+)^{2\nu-1}x_1\,,\qquad
c=(x^+)^{2\nu}\,.
\end{equation}
The deformed background takes the form
\begin{equation}
\begin{aligned}
\tilde{ds}^2&=2dx^+\Big(dx^-+\eta^2\frac{ac}{1+\eta^2c^2}dx^2-\eta^2\frac{bc}{1+\eta^2c^2}dx^1\Big)\\
&-\Big(\frac{k}{(x^+)^2}x_m^2+\eta^2\frac{a^2+b^2}{1+\eta^2c^2}\Big)(dx^+)^2
+\frac{dx_1^2+dx_2^2}{1+\eta^2c^2}
+dx_{m'}^2\,.
%
%
%
%
%
%
\end{aligned}
\end{equation}
With the B-field and dilaton given by
\begin{equation}
\tilde B=-\frac{\eta}{1+\eta^2c^2}\left[(adx^1+bdx^2)\wedge dx^++cdx^2\wedge dx^1\right]\,,\qquad
%
\tilde\Phi=\Phi-\frac12\ln(1+\eta^2c^2)\,.
\end{equation}
One sees from this that
\begin{equation}
\tilde H=4\eta\nu(x^+)^{2\nu-1}dx^2\wedge dx^1\wedge dx^+\,,
\end{equation}
which is independent of the parameter $s$. The fact that also $\Phi$ is independent of $s$ suggests that it might be possible to remove the $s$ dependence also from the metric. Consider the change of coordinates $x_2\rightarrow x_2+f$ and $x^-\rightarrow x^-+gx_2+h$ where $f,g,h$ are functions only of $x^+$. One finds that the choice
\begin{align}
&f=\frac{s}{2}\eta^2(2\nu-1)(x^+)^{\nu+2}\,,\quad
g=-\frac{s}{2}\eta^2(2\nu-1)\nu(x^+)^{\nu+1}\,,
\nonumber\\
&h=\frac{s^2}{8}\eta^2(2\nu-1)^2\left[4(3-2\nu)^{-1}(x^+)^{3-2\nu}-\eta^2\nu(x^+)^{3+2\nu}\right]\,,
%
%
%
\end{align}
removes the dependence on $s$ completely and reduces the background to the one obtained by the TsT with
\begin{equation}
\Theta=k_1\wedge k_4\,.
\end{equation}
Explicitly, the metric is
\begin{equation}
\begin{aligned}
\tilde{ds}^2&=2dx^+\Big(dx^-+\nu\eta^2c^2(1+\eta^2c^2)^{-1}(x_1dx_1+x_2dx^2)/x^+\Big)
\\
&-(x^+)^{-2}\Big(kx_m^2+\nu^2\eta^2c^2(1+\eta^2c^2)^{-1}(x_1^2+x_2^2)\Big)(dx^+)^2
+\frac{dx_1^2+dx_2^2}{1+\eta^2c^2}
+dx_{m'}^2\,.
\end{aligned}
\end{equation}
From (\ref{eq:corrections}) we find the only correction to the deformed background is given by
{
\begin{equation}
\delta G_{++}=
-4\eta^2(2\nu^2-\nu)(x^+)^{4\nu-2}\,,
\end{equation}
which can be canceled by a diffeomorphism $\delta G_{++}=\nabla_+v_+$. In fact the change of coordinates $x^-\rightarrow x^-+\nu\eta^2c^2\frac{x_1^2+x_2^2}{2x^+(1+\eta^2c^2)}$, $x_{1,2}\rightarrow\sqrt{1+\eta^2c^2}\,x_{1,2}$ brings the deformed metric to the form
\begin{equation}
\tilde{ds}^2=
2dx^+dx^-
+(x^+)^{-2}
\left[
-kx_m^2
+\eta^2c^2[-3\nu+5\nu^2-k\eta^2c^2]\frac{x_1^2+x_2^2}{1+\eta^2c^2}
\right]
(dx^+)^2
+dx_m^2\,.
%
%
%
%
%
\end{equation}
Therefore this background is exact at two loops, as is easily checked directly, and possibly to all loops.
}

\subsection{Bianchi type II background}
Next we consider the Bianchi type II background \cite{Gasperini:1994du,Batakis:1995kn} (the $\alpha'$-corrections to Bianchi type I were considered in \cite{Naderi:2017vls})
\begin{equation}
\label{eq:BianchiII}
ds^2=-\cosh(\tau)e^{(a+b+c)\tau}d\tau^2+\frac{e^{a\tau}}{\cosh(\tau)}(dx-zdy)^2+\cosh(\tau)e^{(a+b)\tau}dy^2+\cosh(\tau)e^{(a+c)\tau}dz^2\,,
%
%
%
%
%
%
\end{equation}
supported by a dilaton linear in $\tau$
\begin{equation}
\Phi=a\tau/2\,.
\end{equation}
This solves the Einstein equations provided that the parameters $a,b,c$ are related as
\begin{equation}
bc=a^2+1\,.
\end{equation}
The solution has three Killing vectors
\begin{equation}
k_1=-\partial_z-y\partial_x\,,\quad
k_2=\partial_y\,,\quad
k_3=\partial_x\,,
\end{equation}
which again satisfy a Heisenberg algebra $[k_1,k_2]=k_3$.

From now on we will simplify things by taking $a=0$ and $b=c=1$. The two-loop equations are not automatically satisfied, and we need to find $\alpha'$--corrections for this background. It is convenient to introduce a new coordinate system $\{v,x,y,z\}$ where $v=e^\tau$, since the metric then has a rational dependence on $v$
\be\label{eq:metric-BII}
ds^2=\frac{2 v (dx-z dy)^2}{v^2+1}+\frac{\left(v^2+1\right) \left(v\left(dy^2+dz^2\right)-dv^2\right)}{2 v}\,.
\ee
We assume that the correction to the metric $\delta G_{ij}$ respects the isometries of the background. We turn on the diagonal components $\delta G_{ii}$ and $\delta G_{12}=-z \delta G_{11}$. We also allow for a correction to the dilaton $\delta \Phi$ that, together with $\delta G_{ii}$, is allowed to depend only on $v$. The two-loop equation for the $B$-field is already satisfied. First it is simpler to solve the two-loop equation for the dilaton, because there only the correction $\delta\Phi$ contributes. One finds a second order differential equation $-3 v^6+45 v^4-45 v^2+3-\left(v^2+1\right)^5 \left(v \delta \Phi ''(v)+\delta \Phi'(v)\right)=0$ solved by
\be
\delta \Phi=\frac{v}{2 \left(v^2+1\right)}+\frac{2v}{\left(v^2+1\right)^3}+\tfrac12\arctan{v}+c_{\Phi} \log v\,,
\label{eq:deltaPhi-BII}
\ee
where $c_{\Phi}$ is a constant. Looking at the two-loop equations for the metric, one can find a linear combination of those equations that gives an algebraic constraint imposing $\delta G_{11}=0$. To find $\delta G_{00}, \delta G_{22}, \delta G_{33}$, we first identify linear combinations of the equations that give first order differential equations for $\delta G_{00}$ and $\delta G_{33}$, and we solve them obtaining results written in terms of $\delta G_{22}$. These are then used to get a third order differential equation for $\delta G_{22}$ only, that we also solve. The final result is
\be\label{eq:corr-BII}
\begin{aligned}
\delta G_{00}=&
\frac{-2 v^8+20 v^4+8 v^2-2 \left(v^2-3\right) \left(v^2+1\right)^3 v \arctan{v}+6}{\left(v^4-1\right)^2}\\
&+\frac{\left(v^2+1\right) \left(c_{00} \left(v^2-1\right)^2+v^2 (c_{22}-2
   f_{22})+c_{22}-2 f_{22}-4 c_\Phi \left(v^2-3\right)
   v^2 \log v+8 c_\Phi\right)}{v \left(v^2-1\right)^2}
\,,\\
\delta G_{22}=&
\frac{\left(3 v^2-1\right) \left(\left(v^2+1\right)^3 \arctan{v}+v \left(v^4+2
   v^2+5\right)\right)}{2 \left(v^2-1\right) \left(v^2+1\right)^2}\\
&+\frac{\left(v^2+1\right) \left(v^2 (d_{22}-c_{22})+3
   c_{22}-d_{22}+2 \log v \left(f_{22} \left(v^2-1\right)+4
   c_\Phi\right)-4 f_{22}+8 c_\Phi\right)}{4
   \left(v^2-1\right)}
\,,\\
\delta G_{33}=&
\delta G_{22}-\frac{1}{2} \left(v^2+1\right) (2 c_{00}-2 c_{22}+d_{22}+2
   (f_{22}-6 c_\Phi) \log v+2 f_{22})\,.
\end{aligned}
\ee
For simplicity in what follows we will set all integration constants $c_\Phi= c_{00}= c_{22}= d_{22}=f_{22}=0$. This background admits  a non-abelian deformation with
\begin{equation}
\Theta=\alpha k_1\wedge k_4+\beta k_2\wedge k_3\,,
\end{equation}
where $\alpha,\beta$ are parameters and we have introduced an additional flat direction $w$ so that we can have a fourth Killing vector $k_4=\partial_{w}$. If both $\alpha$ and $\beta$ are non-zero, they can be reabsorbed by redefining $w$ and the deformation parameter $\eta$. For simplicity we set $\alpha=0$, $\beta=1$ and analyze the abelian deformation given by
\begin{equation}
\Theta=k_2\wedge k_3\,.
\end{equation}
The Yang-Baxter deformation to lowest order in $\alpha'$ yields the following deformed background\footnote{We remind that in this paper we use the convention $B=\tfrac{1}{2} B_{ij}dx^i\wedge dx^j$.}
\be\label{eq:Bianchi-k2k3-a0}
\begin{aligned}
ds^2&=\frac{\left(\left(v^2+1\right)^2+4 v z^2\right) dy^2-8 v z dx dy+4 v dx^2}{2 \left(v^2+1\right)
   \left(1+\eta ^2 v\right)}-\frac{\left(v^2+1\right) dv^2}{2 v}+\tfrac12\left(v^2+1\right) dz^2\,,\\
B&=\frac{\eta  v dx\wedge dy}{1+\eta ^2 v}\,,\\
\Phi&=-\frac{1}{2} \log \left(1+\eta ^2 v\right)\,.
\end{aligned}\ee

We can obtain the first $\alpha'$--correction exactly in the deformation parameter $\eta$ if we follow the strategy outlined in section~\ref{sec:Tdual}. The deformation generated by $\Theta=k_2\wedge k_3$ is equivalent to doing first a T-duality along $x$, then shifting $y\to y-\eta \tilde x$ where $\tilde x$ is the dual coordinate to $x$, and then T-dualising $\tilde x$ back.

We first start from the background given by the metric~\eqref{eq:metric-BII} and the $\alpha'$--corrections~\eqref{eq:corr-BII}. This background solves the two-loop equations in the HT scheme, and we need to apply~\eqref{eq:HT-KM} in order to find a solution in the KM scheme. Obviously, since the corrections in~\eqref{eq:HT-KM} are multiplied by an explicit power of $\alpha'$, it is enough to use the uncorrected background to derive them, which simplifies the calculation. Because $B=0$, we can in principle get a non-trivial modification only for the metric from the Ricci tensor, and  for the dilaton from the Ricci scalar. But the Bianchi II background is also Ricci-flat, therefore it is the same in the KM scheme and in the HT scheme.
The next step is that of identifying the fields of the dimensional reduction as in~\eqref{eq:dim-red}. Because we want to do T-duality along $x$ here, we are taking $\xt=x$. This is a straightforward exercise, and instead of writing down all fields of the dimensional reduction, we only write those that can potentially change under the corrected T-duality rules
\be
\sigma=\tfrac12 \log \left(\frac{2v}{1+v^2}\right)\,,\qquad
V=-zdy\,,\qquad
W=0\,,\qquad
b=0\,.
\ee
These particular fields of the dimensional reduction happen not to depend on $\alpha'$ in this particular example. We then implement the $\alpha'$--corrected T-duality rules of KM as in~\eqref{eq:T-KM} and obtain the fields of the dimensional reduction after T-duality
\be
\sigma=-\tfrac12 \log \left(\frac{2v}{1+v^2}\right)
-\alpha'\frac{ \left(v^4-6
   v^2+1\right)}{2 v \left(v^2+1\right)^3}\,,\qquad
V=0\,,\qquad
W=-zdy\,,\qquad
b=0\,.
\ee
After T-duality the scalar $\sigma$ does depend explicitly on $\alpha'$. The explicit form of the two-loop background after performing this first T-duality along $x$ is
\be
\begin{aligned}
ds^2&=
\frac{1}{2}
   \left(v+v^{-1}-\frac{\alpha' \left(v^4-6
   v^2+1\right)}{\left(v^3+v\right)^2}\right) d\tilde x^2\\
&+\frac{1}{2}\left(1+v^2+\frac{\alpha' \left(3 v^2-1\right)
   \left(\left(v^2+1\right)^3 \arctan{v}+v \left(v^4+2
   v^2+5\right)\right)}{\left(v^2-1\right) \left(v^2+1\right)^2}\right)( dy^2 +dz^2)\\
&+
   \left(-\frac{v^2+1}{2 v}-\frac{2 \alpha' \left(v^8-10 v^4-4
   v^2+\left(v^2-3\right) \left(v^2+1\right)^3 v \arctan{v}-3\right)}{\left(v^4-1\right)^2}\right)dv^2\,,\\
B&=zd\tilde x\wedge dy\,,\\
\Phi=&
-\frac{1}{2} \log \left(\frac{2 v}{v^2+1}\right)
+\alpha'\left[\frac{
   \left(2 v^6+3 v^4+16 v^2-1\right)}{4v \left(v^2+1\right)^3}
+\frac{1}{2}  \arctan{v}\right]\,.
\end{aligned}\ee
In the T-dual frame the metric is diagonal (even to two loops) at the cost of having a non-vanishing $B$-field.
We can now do the shift $y\to y-\eta \tilde x$, that here will have only the effect of modifying the metric. To perform  another T-duality along $\tilde x$ we have to first repeat the identification of the fields of the dimensional reduction. We find in particular
\be
\begin{aligned}
\sigma=&
\tfrac{1}{2} \log \left(\tfrac12\left(\eta ^2
   \left(v^2+1\right)+v+v^{-1}\right)\right)\\
&+\tfrac{1}{2} \alpha'\left[-\frac{ \left(v^4-6
   v^2+1\right)}{v \left(v^2+1\right)^3 \left(1+\eta ^2 v\right)}+\frac{ \eta ^2 v \left(3 v^2-1\right) \left(\left(v^2+1\right)^3 \arctan{v}+v
   \left(v^4+2 v^2+5\right)\right)}{\left(v^2-1\right) \left(v^2+1\right)^3 \left(\eta
   ^2 v+1\right)}\right]\,,\\
V&=
\frac{-\eta \ dy}{\left(1+\eta ^2 v\right)^2} \left(v
   \left(1+\eta ^2 v\right)+\frac{\alpha' \left(v \left(3 v^2-1\right)
   \left(v^2+1\right)^2 \arctan{v}+3
   \left(v^6+v^4+v^2\right)-1\right)}{\left(v^2-1\right) \left(v^2+1\right)^2}\right)\,,\\
W&=-zdy\,,\qquad\qquad\qquad\qquad
b=0\,.
\end{aligned}
\ee
At this point we can use again the T-duality rules of KM~\eqref{eq:T-KM}. After doing that we obtain the following background
{
\begingroup
\allowdisplaybreaks
\begin{align}
ds^2=&
-\frac{\left(v^2+1\right) dv^2}{2v}
+\frac{2 (dx-z dy)^2}{\eta ^2\left(v^2+1\right)+v+v^{-1}}
+\frac{\left(v^2+1\right) dy^2}{2(1+\eta ^2 v)}
+\tfrac{1}{2} \left(v^2+1\right) dz^2
\nonumber\\
&
+\alpha'\delta G_{00} dv^2
-4\alpha'\eta^2v^2
\left(
\frac{\delta G_{22}}{\left(v^2+1\right)^2\left(1+\eta ^2 v\right)^2}
+2v\frac{v^2-1}{\left(v^2+1\right)^4 \left(1+\eta ^2 v\right)^2}
\right)(dx-z dy)^2
\nonumber\\
&
+\alpha'\left(\frac{\delta G_{22}}{\left(1+\eta ^2 v\right)^2}-\eta^2\frac{v^4-6v^2+1}{2\left(v^2+1\right)^2\left(1+\eta ^2 v\right)^2}\right)dy^2
+\alpha'\delta G_{22}dz^2
\,,\nonumber\\
\tilde B&=
\frac{\alpha' \eta dv\wedge dz}{\left(v^2+1\right) \left(1+\eta ^2 v\right)}\nonumber\\
&
+\eta  v dx\wedge dy
\Big(
\frac{1}{1+\eta ^2 v}
+2\alpha'\frac{\delta G_{22}}{\left(v^2+1\right)\left(1+\eta ^2 v\right)^2}
+\alpha'
\frac{
2v(v^2-3)
+\eta ^2\left(3 v^2-1\right)\left(v^2-1\right)
}{\left(v^2+1\right)^3\left(1+\eta ^2 v\right)^3}
\Big)
\,,\nonumber\\
\tilde\Phi&=-\tfrac12 \log \left(1+\eta ^2v\right)
+\alpha'\delta\Phi
-\alpha'\eta^2\frac{4v(v^2+1)^2\delta G_{22}+5v^4-10v^2+1}{4\left(v^2+1\right)^3\left(1+\eta ^2 v\right)}
\,,
\end{align}
where $\delta G_{ij}$ and $\delta\Phi$ are the corrections to the undeformed background given in (\ref{eq:deltaPhi-BII}) and (\ref{eq:corr-BII}). This is a TsT of the initial Bianchi II that solves the two-loop equations in the KM scheme. To go to the HT scheme we use again~\eqref{eq:HT-KM}. Because of the deformation, now the dictionary  to go to the new scheme is non-trivial, and the background in the HT scheme reads
\be
\label{eq:BII-corr1}
\begin{aligned}
ds^2=&
\tilde G_{ij}dx^idx^j\,,\\
\tilde B&=\frac{\alpha' \eta  dv\wedge dz}{\left(v^2+1\right) \left(1+\eta ^2 v\right)}\\
&{}
+\eta  v dx\wedge dy
\left(
\frac{1}{1+\eta ^2 v}
+2\alpha'\frac{\delta G_{22}}{\left(v^2+1\right)\left(1+\eta ^2 v\right)^2}
+2\alpha'v\frac{v^2-3}{\left(v^2+1\right)^3\left(1+\eta ^2 v\right)^2}
\right)
\,,\\
\tilde\Phi&=-\tfrac12\log \left(1+\eta ^2v\right)
+\alpha'\delta\Phi
-\alpha'\eta^2\frac{v\delta G_{22}}{\left(v^2+1\right)\left(1+\eta ^2 v\right)}
-\alpha'\eta^2\frac{3v^4-14v^2-1}{4\left(v^2+1\right)^3 \left(1+\eta ^2 v\right)}
\,,
\end{aligned}
\ee
where
\begin{align}
\tilde G_{00}&=-\frac{v^2+1}{2 v}
+\alpha'\delta G_{00}
-\alpha'\eta ^2\frac{\eta ^2+3 \eta ^2 v^2+2 v}{2\left(v^2+1\right) \left(1+\eta ^2 v\right)^2}
\,,\nonumber\\
\tilde G_{11}&=
\frac{2v}{\left(v^2+1\right)\left(1+\eta ^2 v\right)}
-4\alpha'\eta^2v^2\frac{\delta G_{22}}{\left(v^2+1\right)^2\left(1+\eta ^2 v\right)^2}
-4v\alpha'\eta^2v^2\frac{v^2-3}{\left(v^2+1\right)^4\left(1+\eta ^2 v\right)^2}
\,,\nonumber\\
\tilde G_{22}&=
\frac{v^2+1}{2\left(1+\eta ^2 v\right)}
-\alpha'\eta^2v^2\frac{v^2-3}{\left(v^2+1\right)^2\left(1+\eta ^2 v\right)^2}
+\alpha'\frac{\delta G_{22}}{\left(1+\eta ^2 v\right)^2}
+z^2\tilde G_{11}
\,,\nonumber\\
\tilde G_{33}&=
\tfrac12\left(v^2+1\right)
+\alpha'\delta G_{22}
-\alpha'\eta^2\frac{v^2}{\left(v^2+1\right)\left(1+\eta ^2v\right)}
\,,\nonumber\\
\tilde G_{12}&={-z\tilde G_{11}}\,.
\label{eq:BII-corr2}
\end{align}
\endgroup
Performing the redefinition of the dilaton given in (\ref{eq:dilaton-scheme}) this background agrees precisely with that obtained from the all order expression (\ref{eq:corrections-all-orders}).
}

\vspace{12pt}

When we want to work out a deformation generated by $\Theta=k_1\wedge k_4$ following the strategy of section~\ref{sec:Tdual}, we first need to find a coordinate system in which $k_1$ acts as a simple shift of a coordinate. We can redefine
\be
x=x'+y'z'\,,\qquad
y=y'\,,\qquad
z=z'\,,
\ee
so that in the new coordinate system $k_1=-\partial_{z'}$. 
As should be clear from the discussion at the beginning of this section, the isometry generated by $k_1$ is not broken by $\alpha'$ corrections, therefore the metric will not depend on $z'$ also at two loops. The deformation generated by $\Theta=k_1\wedge k_4$ can be obtained by doing T-duality $w\to \tilde w$, then the shift $z'\to z'-\eta \tilde w$, and then T-duality back $\tilde w\to w$. We will omit the explicit results for this particular deformation, since they involve very long expressions, and we have already presented our method in the previous deformation generated by $\Theta=k_2\wedge k_3$. {We have checked that the resulting background again agrees with that obtained by the $\alpha'$-corrected open-closed string map (\ref{eq:corrections-all-orders}).}

The interesting point is that we can combine these two TsT transformations. We can first do a TsT involving $x$ and $y$ corresponding to $\Theta=k_2\wedge k_3$. At the end of this result the background is still invariant under isometries generated by $k_1$ and $k_4$, and we can do a second TsT transformation involving $z'$ and $w$,  equivalent to $\Theta=k_1\wedge k_4$. The composition of the two deformations is equivalent to the deformation given by $\Theta=k_1\wedge k_4+k_2\wedge k_3$, as explained in~\cite{Borsato:2018idb}. The non-abelian nature of the deformation is related to the fact that if we had started from $\Theta=k_1\wedge k_4$ instead, we would have broken the isometries that we would need to perform the deformation with $\Theta=k_2\wedge k_3$. As follows from the results of~\cite{Borsato:2018idb}, in the maximally deformed limit $\eta\to \infty$ we recover the non-abelian T-dual of the original Bianchi II solution, where the isometries dualized are those corresponding to the Killing vectors $k_1,k_2,k_3$ forming a Heisenberg algebra, and $k_4$. By this argument it follows that non-abelian T-dual models related to this class of Yang-Baxter deformations remain conformal on the worldsheet to two loops. Because T-duality remains a symmetry of the string at higher orders in an $\alpha'$-expansion, we can argue that this is true to all loops. {Unfortunately the all order expression (\ref{eq:corrections-all-orders}) turns out not to give the correct answer in this case.}

\section{Conclusions}
We have argued that (homogeneous) YB deformed string $\sigma$--models that are conformal at one loop remain conformal at two loops,\footnote{Provided, of course, the undeformed background is conformal to two loops.} i.e. including the first correction in $\alpha'$. We showed this to second order in the deformation parameter $\eta$ for a generic unimodular deformation of a background with vanishing $B$-field. We also argued that using the $\alpha'$--corrected T-duality rules of \cite{Kaloper:1997ux} one can verify this to all orders in the deformation parameter for the cases that can be built from TsT transformations, and we explained that this strategy can be used also for the non-abelian YB deformations that are equivalent to a non-commuting sequence of TsT transformations\footnote{See e.g. \cite{Borsato:2016ose,vanTongeren:2016eeb}}. We exemplified our results in the case of a deformation of a Bianchi type II background.

Our findings suggest that one-loop conformal YB $\sigma$--models should in fact remain conformal to first order in $\alpha'$, and likely all orders. Since these models can be thought of as a generalization of non-abelian T-duality~\cite{Hoare:2016wsk,Borsato:2016pas,Borsato:2018idb} (which can be recovered in an appropriate $\eta\to \infty$ limit) our findings suggest that the same should be true for NATD. This was also argued recently from a different perspective in \cite{Hoare:2019ark,Hoare:2019mcc}, studying renormalizability of a different type of integrable deformation of $\sigma$--models.\footnote{Early work on $\alpha'$--corrections in NATD include \cite{Subbotin:1995zz,Balog:1996im,Balazs:1997be,Bonneau:2001za}.} To test this idea one should start from a model which is conformal to all orders in $\alpha'$ and then deform it. A good candidate is therefore the unimodular deformation of $AdS_3\times S^3$ constructed in \cite{Borsato:2018spz}.

{
We saw that the expression (\ref{eq:corrections-all-orders}) for the all order in $\eta$ form of the first $\alpha'$-correction to YB deformations works in simple cases but fails in general. It is an important problem to fix it so that it holds in general. If a simple solution exists for the corrections, it is also interesting in the special case of TsT transformations, whose corrections have, to our knowledge, not been analyzed before. If, further more, this continues to work to higher orders in $\alpha'$ it could even help in determining the structure of higher $\alpha'$--corrections to the target space equations of motion. This approach could be said to be an example of using $O(d,d)$ symmetry to determine/constrain higher $\alpha'$--corrections.
}

We plan to address some of these questions in the near future.

\section*{Acknowledgements}
We thank B. Hoare, N. Levine and A. Tseytlin for useful and interesting discussions. LW wishes to thank the participants of the workshop ''New  frontiers  of  integrable  deformations'' in Villa Garbald, Castasegna for interesting discussions. RB is grateful to J. Edelstein, J. A. Sierra-Garcia, and in particular to A. Vilar L\'opez for very useful discussions. The work of RB is supported by the fellowship of ``la Caixa Foundation'' (ID 100010434) with code LCF/BQ/PI19/11690019. He is also supported by the Maria de Maeztu Unit of Excellence MDM-2016-0692, by FPA2017-84436-P, by Xunta de Galicia (ED431C 2017/07), and by FEDER.

\appendix

\section{Killing identities}
The Killing vectors satisfy the equations (suppressing the Lie algebra index)
\begin{equation}
\nabla_{(i}k_{j)}=0\qquad\nabla_i\nabla_jk_l=R_{ljin}k^n\,.
\end{equation}
Using this and the expression for $\Theta$ in (\ref{eq:Theta}) we can derive the useful two-derivative identity
\begin{equation}
2\nabla_k\nabla_{(i}\Theta_{j)l}
=
2\nabla_kk_{(i}\times\nabla_{j)}k_l
+2R_{knl(i}\Theta_{j)}{}^n
=
-\nabla_{(i}\nabla_{j)}\Theta_{kl}
+2R_{knl(i}\Theta_{j)}{}^n
-R_{k(ij)n}\Theta_l{}^n
+R_{l(ij)n}\Theta_k{}^n\,.
\label{eq:2-der-id}
\end{equation}
A special case of this is
\begin{equation}
\nabla^2\Theta_{ij}
=
-R_{ijkl}\Theta^{kl}
+R_{ik}\Theta_j{}^k
-R_{jk}\Theta_i{}^k\,.
\end{equation}

In addition we have the unimodularity condition, which in terms of $\Theta$, takes the form
\begin{equation}
\nabla_k\Theta^{kl}=0\,.
\label{eq:unimodular}
\end{equation}

We also know that the dilaton respects the isometries so that
\begin{equation}
k^i\nabla_i\Phi=0\,.
\label{eq:dilaton}
\end{equation}

Using these facts we can prove the useful identity
\begin{equation}
\nabla_k(R_{ijlm}\Theta^{lm})
=
-\tfrac12R_{iklm}\nabla_j\Theta^{lm}
+R_{imkl}\nabla^m\Theta_j{}^l
-R_{ilmk}\nabla^m\Theta_j{}^l
-(i\leftrightarrow j)\,.
\label{eq:id0}
\end{equation}
This follows by noting that
\begin{align}
2R_{ijlm}\nabla_k\Theta^{lm}
=&
-4\nabla_m\nabla_jk_i\times\nabla_kk^m
=
-4\nabla_m(\nabla_jk_i\times\nabla_kk^m)
+4R_{kl}\nabla_jk_i\times k^l
\nonumber\\
=&
-2\nabla_m\nabla_j(k_i\times\nabla_kk^m)
+2\nabla_m(R_{mkjl}\Theta_i{}^l)
+2R_{kl}\nabla_jk_i\times k^l
-(i\leftrightarrow j)
\nonumber\\
=&
-\nabla^m\nabla_j\nabla_k\Theta_{im}
+\nabla^m\nabla_j\nabla_m\Theta_{ik}
+\nabla^m\nabla_j\nabla_i\Theta_{mk}
+2\nabla^m(R_{mkjl}\Theta_i{}^l)
\nonumber\\
&{}
+2R_{kl}\nabla_jk_i\times k^l
-(i\leftrightarrow j)
\nonumber\\
=&
\tfrac12R_{ijlm}\nabla_k\Theta^{lm}%
-\tfrac12\nabla_kR_{ijlm}\Theta^{lm}%
-\tfrac12R_{iklm}\nabla_j\Theta^{lm}%
+R_{imkl}\nabla^m\Theta_j{}^l%
\nonumber\\
&{}
-R_{ilmk}\nabla^m\Theta_j{}^l%
-(i\leftrightarrow j)\,,
\end{align}
where we have used the fact that 
\begin{equation}
\nabla^l\Phi\,\nabla_k\Theta_{lj}
+\nabla^l\Phi\,\nabla_l\Theta_{kj}
+\nabla^l\Phi\,\nabla_j\Theta_{kl}=0\,,
\end{equation}
as is easily verified. Acting with $\nabla^k$, and using also $\nabla^k\Phi$ times the above identity, one finds
\begin{align}
4\nabla^k(R_{ijlm}\nabla_k\Theta^{lm})
=&
3\nabla^k\nabla_{[i}(R_{jk]lm}\Theta^{lm})
-2R_{imkl}R_{jn}{}^{kl}\Theta^{mn}
+4R_i{}^{klm}R_{jkln}\Theta_m{}^n
\nonumber\\
&{}
+2R_{ijlm}\nabla^k\Phi\,\nabla_k\Theta^{lm}
+4R_{iklm}\nabla^k\Phi\,\nabla_j\Theta^{lm}
-(i\leftrightarrow j)\,.
\label{eq:id1}
\end{align}

\section{Relations needed for second order calculation}\label{app:relations}
For the second order calculations we define the following `basis' of terms (for readability we write all indices as lower indices)
\begingroup
\allowdisplaybreaks
\begin{align}
f_1=&R_{ilmn}\nabla_j\Theta_{mn}\Theta_{kl} & f_{12}=&R_{imkn}\nabla_m\Theta_{lj}\Theta_{ln} & f_{23}=&R_{klmn}\nabla_m\Theta_{in}\Theta_{jl}\nonumber\\
f_2=&R_{ilmn}\nabla_j\Theta_{kl}\Theta_{mn} & f_{13}=&R_{imkn}\nabla_l\Theta_{mj}\Theta_{ln} & f_{24}=&R_{klmn}\nabla_l\Theta_{im}\Theta_{jn}\nonumber\\
f_3=&R_{ikmn}\nabla_j\Theta_{ml}\Theta_{ln} & f_{14}=&R_{ilmn}\nabla_k\Theta_{mn}\Theta_{lj} & f_{25}=&\nabla_lR_{ikmn}\Theta_{jl}\Theta_{mn}\nonumber\\
f_4=&R_{imnk}\nabla_j\Theta_{ml}\Theta_{ln} & f_{15}=&R_{ilmn}\nabla_l\Theta_{mn}\Theta_{kj} & f_{26}=&\nabla_kR_{ilmn}\Theta_{jl}\Theta_{mn}\nonumber\\
f_5=&R_{ilmn}\nabla_l\Theta_{mj}\Theta_{nk} & f_{16}=&R_{ilmn}\nabla_l\Theta_{km}\Theta_{nj} & f_{27}=&\nabla_i\nabla_j\Theta_{mn}\nabla_k\Theta_{mn}\nonumber\\
f_6=&R_{ilmn}\nabla_m\Theta_{lj}\Theta_{nk} & f_{17}=&R_{ilmn}\nabla_m\Theta_{kn}\Theta_{lj} & f_{28}=&\nabla_i\nabla_k\Theta_{mn}\nabla_j\Theta_{mn}\\
f_7=&R_{ilmn}\nabla_m\Theta_{nj}\Theta_{lk} & f_{18}=&R_{ikmn}\nabla_m\Theta_{ln}\Theta_{lj} & f_{29}=&\nabla_i\nabla_j\Theta_{mn}\nabla_m\Theta_{nk}\nonumber\\
f_8=&R_{ilmn}\nabla_l\Theta_{kj}\Theta_{mn} & f_{19}=&R_{ikmn}\nabla_l\Theta_{mn}\Theta_{lj} & f_{30}=&\nabla_i\nabla_k\Theta_{mn}\nabla_m\Theta_{nj}\nonumber\\
f_9=&R_{ilmn}\nabla_k\Theta_{lj}\Theta_{mn} & f_{20}=&R_{imkn}\nabla_m\Theta_{ln}\Theta_{lj} & f_{31}=&\nabla_i\nabla_m\Theta_{nk}\nabla_j\Theta_{mn}\nonumber\\
f_{10}=&R_{ikmn}\nabla_m\Theta_{lj}\Theta_{ln} & f_{21}=&R_{klmn}\nabla_i\Theta_{jl}\Theta_{mn} & f_{32}=&\nabla_i\nabla_m\Theta_{nk}\nabla_m\Theta_{nj}\nonumber\\
f_{11}=&R_{ikmn}\nabla_l\Theta_{mj}\Theta_{ln} & f_{22}=&R_{klmn}\nabla_i\Theta_{mn}\Theta_{jl} & f_{33}=&\nabla_i\nabla_m\Theta_{nk}\nabla_n\Theta_{mj}\nonumber
\end{align}
where we suppress the free indices $ijk$ and assume symmetry in $ij$ throughout. We also define the terms with only one free index
\begin{align}
\hat f_1=&R_{klmn}\nabla_j\Theta_{kl}\Theta_{mn} & \hat f_4=&R_{jlmn}\nabla_k\Theta_{mn}\Theta_{kl} & \hat f_7=&\nabla_j\nabla_l\Theta_{mn}\nabla_l\Theta_{mn}\nonumber\\
\hat f_2=&R_{klmn}\nabla_k\Theta_{lj}\Theta_{mn} & \hat f_5=&R_{jlmn}\nabla_m\Theta_{nk}\Theta_{kl} & \hat f_8=&\nabla_j\nabla_l\Theta_{mn}\nabla_m\Theta_{nl}\\
\hat f_3=&R_{klmn}\nabla_m\Theta_{kl}\Theta_{nj} & \hat f_6=&R_{jlmn}\nabla_l\Theta_{km}\Theta_{kn}\nonumber
\end{align}

We will denote for example $\nabla^kf_{1ijk}$ as $\nabla\cdot f_1$, again suppressing the indices, and similarly for example $\nabla_{(i}\hat f_{1j)}$ as $\nabla\hat f_1$. Using the Killing vector identities, unimodularity and isometry of the dilaton one finds
\begin{align}
\nabla\cdot f_1=&{}\tfrac12g_{12}+g_{23}-2h_6
\\
\nabla\cdot f_2=&{}\tfrac12g_{12}+g_{15}-\tfrac12h_1+2m_1
\\
\nabla\cdot f_3=&g_{13}-g_{25}-h_5-h_7-2m_5-2m_6
\\
\nabla\cdot f_4=&g_{14}-g_{24}+g_{25}+\tfrac12h_5+h_7-\tfrac12h_8+2m_6
\\
\nabla\cdot f_5=&-\tfrac12g_1+\tfrac12g_{23}-g_{25}+\tfrac12h_3-h_5-h_6-\tfrac12h_8
\\
\nabla\cdot f_6=&-\tfrac12g_1-\tfrac12g_{10}-\tfrac12g_{23}+g_{24}-g_{25}-\tfrac12h_3-h_5+\tfrac12h_8
\\
\nabla\cdot f_7=&-\tfrac12g_{10}-g_{23}+g_{24}-h_3+h_6+h_8
\\
\nabla\cdot f_8=&g_1+g_3+\tfrac12h_1+2m_2
\\
\nabla\cdot f_9=&g_1+g_8+g_{10}+h_1+2m_1-2m_2
\\
\nabla\cdot f_{10}=&g_4+g_7+g_{24}-2g_{25}+h_4-\tfrac12h_5-h_6-h_7+\tfrac12h_8+2m_7-2m_8
\\
\nabla\cdot f_{11}=&g_5-\tfrac12g_{23}+g_{25}-h_4+\tfrac12h_5+h_6+\tfrac12h_8+2m_{10}-2m_{11}
\\
\nabla\cdot f_{12}=&g_4+g_{24}-2g_{25}-\tfrac12h_3+h_4-\tfrac12h_5-h_6-h_7+2m_7
\\
\nabla\cdot f_{13}=&g_6+\tfrac12g_{23}-g_{24}+g_{25}+\tfrac12h_3-h_4+h_5-h_6+\tfrac12h_8+2m_{10}
\\
\nabla\cdot f_{14}=&g_2+g_8+h_2+2m_3
\\
\nabla\cdot f_{15}=&g_2-g_{11}+g_{21}
\\
\nabla\cdot f_{16}=&-\tfrac12g_2-g_5+\tfrac12g_{11}-\tfrac14h_2+2m_{16}
\\
\nabla\cdot f_{17}=&\tfrac12g_2+g_6+\tfrac12h_2-m_3
\\
\nabla\cdot f_{18}=&g_4-\tfrac32g_{21}+h_3+h_4-h_{10}+2m_9+2m_{17}
\\
\nabla\cdot f_{19}=&g_3+g_{21}-2h_4+2m_4
\\
\nabla\cdot f_{20}=&g_4+g_7-\tfrac32g_{21}+\tfrac32h_3+h_4-\tfrac12h_{10}+2m_9
\\
\nabla\cdot f_{21}=&g_9-\tfrac12g_{20}-\tfrac12h_2+h_9+2m_{14}
\\
\nabla\cdot f_{22}=&-g_{16}-g_{22}+2h_4-2m_{13}
\\
\nabla\cdot f_{23}=&g_{17}+\tfrac32g_{22}-h_3-h_4-h_{11}+2m_{12}
\\
\nabla\cdot f_{24}=&-g_{18}-\tfrac12h_3+\tfrac12h_{11}+2m_{18}
\\
\nabla\cdot f_{25}=&g_{11}-g_1-g_2-\tfrac12h_2-2h_4-4m_2+2m_{19}
\\
\nabla\cdot f_{26}=&-g_1-g_2-g_{10}-h_2-4h_4-2m_{15}
\\
\nabla\cdot f_{27}=&2g_5-2g_6+2g_7-2g_{13}-2g_{14}-g_{20}+2g_{28}+2g_{32}+4m_{21}
\\
\nabla\cdot f_{28}=&-g_{12}+2g_{13}-g_{19}-2g_{25}+g_{26}+2m_{20}
\\
\nabla\cdot f_{29}=&\tfrac12g_3-g_4-g_5-g_7+\tfrac12g_8+g_{13}+\tfrac12g_{15}+\tfrac12g_{20}-g_{29}+g_{30}-g_{33}+g_{34}+2m_{21}
\\
\nabla\cdot f_{30}=&g_4-g_5+g_7-g_{10}-g_{16}-g_{22}+g_{23}-g_{25}+g_{26}-g_{27}+2m_{22}
\\
\nabla\cdot f_{31}=&\tfrac12g_{12}-g_{13}-g_{14}+\tfrac12g_{15}+\tfrac12g_{19}-\tfrac12g_{23}-g_{24}+g_{25}+g_{27}+2m_{23}
\\
\nabla\cdot f_{32}=&-g_7+\tfrac12g_8+\tfrac12g_{10}+\tfrac12g_{16}+\tfrac12g_{22}-g_{23}+g_{24}+g_{25}-g_{26}+g_{27}+2m_{24}
\\
\nabla\cdot f_{33}=&\tfrac12g_3-g_5+g_6-\tfrac12g_{10}-\tfrac12g_{16}-\tfrac12g_{22}-\tfrac12g_{23}+2m_{25}
\end{align}
and
\begin{align}
\nabla\hat f_1=&g_{12}+g_{19}+g_{20}
\\
\nabla\hat f_2=&g_{10}+g_{16}+\tfrac12g_{20}+h_2
\\
\nabla\hat f_3=&-g_9+g_{11}+g_{22}
\\
\nabla\hat f_4=&g_{15}+g_{23}-2g_{33}
\\
\nabla\hat f_5=&-g_{14}+g_{24}+g_{32}+g_{34}
\\
\nabla\hat f_6=&g_{13}+g_{25}+g_{34}+g_{35}
\\
\nabla\hat f_7=&-g_4-g_6-3g_{14}+g_{26}+g_{28}-g_{29}-g_{30}+2g_{32}+2g_{34}
\\
\nabla\hat f_8=&
\tfrac12
\left(
g_3
-g_4
-g_6
+g_8
-3g_{14}
+3g_{15}
+2g_{27}
+g_{28}
-g_{29}
+g_{30}
-g_{31}
+2g_{32}
-4g_{33}
+2g_{34}
\right)
\end{align}
where we have defined the $\nabla^2R\Theta^2$-terms
\begin{align}
g_1=&\nabla_kR_{ilmn}\nabla_l\Theta_{kj}\Theta_{mn} & g_{13}=&R_{ilmn}\nabla_j\Theta_{mk}\nabla_l\Theta_{kn} & g_{25}=&R_{ilmn}\nabla_j\nabla_l\Theta_{km}\Theta_{kn}\nonumber\\
g_2=&\nabla_kR_{ilmn}\nabla_k\Theta_{mn}\Theta_{lj} & g_{14}=&R_{ilmn}\nabla_j\Theta_{kl}\nabla_m\Theta_{kn} & g_{26}=&\nabla_i\nabla_k\Theta_{mn}\nabla_j\nabla_k\Theta_{mn}\nonumber\\
g_3=&R_{ilmn}\nabla_l\Theta_{kj}\nabla_k\Theta_{mn} & g_{15}=&R_{ilmn}\nabla_j\Theta_{kl}\nabla_k\Theta_{mn} & g_{27}=&\nabla_i\nabla_k\Theta_{mn}\nabla_j\nabla_m\Theta_{nk}\nonumber\\
g_4=&R_{ilmn}\nabla_l\Theta_{kj}\nabla_m\Theta_{kn} & g_{16}=&R_{klmn}\nabla_i\Theta_{mn}\nabla_k\Theta_{lj} & g_{28}=&R_{kijl}\nabla_m\Theta_{nk}\nabla_m\Theta_{nl}\nonumber\\
g_5=&R_{ilmn}\nabla_k\Theta_{mj}\nabla_l\Theta_{kn} & g_{17}=&R_{klmn}\nabla_k\Theta_{li}\nabla_m\Theta_{nj} & g_{29}=&R_{kijl}\nabla_m\Theta_{nk}\nabla_n\Theta_{ml}\nonumber\\
g_6=&R_{ilmn}\nabla_k\Theta_{lj}\nabla_m\Theta_{kn} & g_{18}=&R_{klmn}\nabla_k\Theta_{mi}\nabla_l\Theta_{nj} & g_{30}=&R_{kijl}\nabla_k\Theta_{mn}\nabla_m\Theta_{nl}\\
g_7=&R_{ilmn}\nabla_m\Theta_{kj}\nabla_n\Theta_{kl} & g_{19}=&R_{klmn}\nabla_i\Theta_{kl}\nabla_j\Theta_{mn} & g_{31}=&R_{kijl}\nabla_k\Theta_{mn}\nabla_l\Theta_{mn}\nonumber\\
g_8=&R_{ilmn}\nabla_k\Theta_{lj}\nabla_k\Theta_{mn} & g_{20}=&R_{klmn}\nabla_i\nabla_j\Theta_{kl}\Theta_{mn} & g_{32}=&\nabla_mR_{kijl}\nabla_m\Theta_{nk}\Theta_{nl}\nonumber\\
g_9=&R_{klmn}\nabla_i\Theta_{jl}\nabla_k\Theta_{mn} & g_{21}=&R_{ilmn}\nabla_k\nabla_l\Theta_{mn}\Theta_{kj} & g_{33}=&\nabla_mR_{kijl}\nabla_n\Theta_{mk}\Theta_{nl}\nonumber\\
g_{10}=&\nabla_iR_{klmn}\nabla_k\Theta_{lj}\Theta_{mn} & g_{22}=&R_{klmn}\nabla_i\nabla_k\Theta_{mn}\Theta_{lj} & g_{34}=&\nabla_mR_{kijl}\nabla_k\Theta_{mn}\Theta_{nl}\nonumber\\
g_{11}=&\nabla_iR_{klmn}\nabla_k\Theta_{mn}\Theta_{lj} & g_{23}=&R_{ilmn}\nabla_j\nabla_k\Theta_{mn}\Theta_{kl} & g_{35}=&\nabla_mR_{kijl}\nabla_k\Theta_{ln}\Theta_{mn}\nonumber\\
g_{12}=&\nabla_iR_{klmn}\nabla_j\Theta_{kl}\Theta_{mn} & g_{24}=&R_{ilmn}\nabla_j\nabla_m\Theta_{nk}\Theta_{kl}\nonumber
\end{align}
the $R^2\Theta^2$-terms
\begin{align}
h_1=&R_{ipkl}R_{jpmn}\Theta_{kl}\Theta_{mn} & h_5=&R_{ilmn}R_{jlkp}\Theta_{mk}\Theta_{np} & h_9=&R_{kijp}R_{klmn}\Theta_{mn}\Theta_{pl}\nonumber\\
h_2=&R_{ilmn}R_{mnkp}\Theta_{kp}\Theta_{jl} & h_6=&R_{ikmp}R_{jlnp}\Theta_{kl}\Theta_{mn} & h_{10}=&R_{klmi}R_{klmn}(\Theta^2)_{nj}\\
h_3=&R_{ilmn}R_{mnkp}\Theta_{kl}\Theta_{jp} & h_7=&R_{ilmn}R_{jlmk}(\Theta^2)_{nk} & h_{11}=&R_{klmn}R_{klmp}\Theta_{in}\Theta_{jp}\nonumber\\
h_4=&R_{ilmn}R_{klmp}\Theta_{np}\Theta_{jk} & h_8=&R_{ilmn}R_{jkmn}(\Theta^2)_{lk}\nonumber
\end{align}
and the terms involving the dilaton
\begin{align}
m_1=&R_{ilmn}\nabla_k\Phi\,\nabla_j\Theta_{kl}\Theta_{mn} & m_{10}=&R_{ilmn}\nabla_m\Phi\,\nabla_k\Theta_{lj}\Theta_{kn} & m_{19}=&\nabla_kR_{ilmn}\nabla_l\Phi\,\Theta_{kj}\Theta_{mn}\nonumber\\
m_2=&R_{ilmn}\nabla_k\Phi\,\nabla_l\Theta_{kj}\Theta_{mn} & m_{11}=&R_{ilmn}\nabla_m\Phi\,\nabla_k\Theta_{nj}\Theta_{kl} & m_{20}=&\nabla_k\Phi\,\nabla_i\nabla_k\Theta_{mn}\nabla_j\Theta_{mn}\nonumber\\
m_3=&R_{ilmn}\nabla_k\Phi\,\nabla_k\Theta_{mn}\Theta_{lj} & m_{12}=&R_{klmn}\nabla_k\Phi\,\nabla_m\Theta_{ni}\Theta_{lj} & m_{21}=&\nabla_k\Phi\,\nabla_i\nabla_j\Theta_{mn}\nabla_m\Theta_{nk}\nonumber\\
m_4=&R_{ilmn}\nabla_l\Phi\,\nabla_k\Theta_{mn}\Theta_{kj} & m_{13}=&R_{klmn}\nabla_k\Phi\,\nabla_i\Theta_{mn}\Theta_{lj} & m_{22}=&\nabla_k\Phi\,\nabla_i\nabla_k\Theta_{mn}\nabla_m\Theta_{nj}\nonumber\\
m_5=&R_{ilmn}\nabla_m\Phi\,\nabla_j\Theta_{nk}\Theta_{kl} & m_{14}=&R_{klmn}\nabla_k\Phi\,\nabla_i\Theta_{jl}\Theta_{mn} & m_{23}=&\nabla_k\Phi\,\nabla_i\nabla_m\Theta_{nk}\nabla_j\Theta_{mn}\nonumber\\
m_6=&R_{ilmn}\nabla_m\Phi\,\nabla_j\Theta_{kl}\Theta_{kn} & m_{15}=&\nabla_kR_{ilmn}\nabla_k\Phi\,\Theta_{mn}\Theta_{lj} & m_{24}=&\nabla_k\Phi\,\nabla_i\nabla_m\Theta_{nk}\nabla_m\Theta_{nj}\nonumber\\
m_7=&R_{ilmn}\nabla_m\Phi\,\nabla_l\Theta_{kj}\Theta_{kn} & m_{16}=&R_{ilmn}\nabla_k\Phi\,\nabla_l\Theta_{km}\Theta_{nj} & m_{25}=&\nabla_k\Phi\,\nabla_i\nabla_m\Theta_{nk}\nabla_n\Theta_{mj}\nonumber\\
m_8=&R_{ilmn}\nabla_m\Phi\,\nabla_n\Theta_{kj}\Theta_{kl} & m_{17}=&R_{ilmn}\nabla_m\Phi\,\nabla_n\Theta_{lk}\Theta_{kj}\nonumber\\
m_9=&R_{ilmn}\nabla_m\Phi\,\nabla_l\Theta_{kn}\Theta_{kj} & m_{18}=&R_{klmn}\nabla_k\Phi\,\nabla_l\Theta_{mi}\Theta_{nj}
\end{align}

\endgroup

\subsection{Additional identities}
Contracting (\ref{eq:id0}) with $\Theta$ and one covariant derivative, or the derivative of the dilaton, in all possible ways gives the identities
\begin{align}
0
=&
g_{12}
-4g_{13}
-2g_{14}
+g_{15}
+g_{19}
\,,
\label{eq:id03}
\\
0=&
2g_1
+g_3
-2g_4
-4g_7
+2g_8
+2g_{10}
+g_{16}
+2g_{17}
-4g_{18}
\,,
\label{eq:id02}
\\
0
=&
2g_1
+2g_3
-4g_5
+2g_6
+g_8
-g_{16}
+2g_{17}
-4g_{18}
\,,
\label{eq:id01}
\\
0=&
\hat f_1
+\hat f_4
+2\hat f_5
-4\hat f_6
+\nabla_iR_{klmn}\Theta_{mn}\Theta_{kl}
\,,
\label{eq:id04}
\\
0=&
2m_4
+2m_{12}
-m_{13}
+4m_{16}
+4m_{18}
+2m_{19}
\,,
\\
0=&
2m_3
+m_4
-2m_9
+2m_{12}
+m_{13}
+2m_{15}
+2m_{17}
+4m_{18}
\,,
\\
0=&
2f_{14}
+2f_{18}
+f_{19}
-4f_{20}
-f_{22}
+2f_{23}
+4f_{24}
-2f_{26}
\,,
\\
0=&
f_{14}
+4f_{16}
+2f_{17}
+2f_{19}
+f_{22}
+2f_{23}
+4f_{24}
-2f_{25}
\,.
\end{align}
The last two imply, using the previous ones, that $m_{19}=m_2$ and
\begin{equation}
0=
g_2
+g_{21}
+g_{22}
+h_2
-2h_3
\label{eq:f14id}
\end{equation}
In addition we can derive the following identity
\begin{align}
2h_5
=&
2R_{iklp}R_{jkmn}\Theta^{pm}\Theta_{nl}
=
-2\nabla_l\nabla_kk_i\times\nabla_n\nabla_kk_j\Theta_{nl}
\nonumber\\
=&
-2\nabla_l(R_{jknm}\nabla_kk_i\times k_m\Theta_{nl})
+R_{lnkm}\nabla_kk_i\times\nabla_mk_j\Theta_{nl}
+R_{lnjm}\nabla_kk_i\times\nabla_kk_m\Theta_{nl}
\nonumber\\
=&
\nabla_l(R_{jknm}\nabla_k\Theta_{mi}\Theta_{nl})
+\nabla_l(R_{jknm}\nabla_m\Theta_{ki}\Theta_{nl})
+\nabla_l(R_{jknm}\nabla_i\Theta_{km}\Theta_{nl})
\nonumber\\
&{}
-\tfrac12R_{klmn}\Theta_{kl}\nabla_i\nabla_j\Theta_{mn}
-R_{lnkm}R_{mijp}\Theta_{kp}\Theta_{nl}
-\tfrac12R_{klpi}R_{jpmn}\Theta_{kl}\Theta_{mn}
\nonumber\\
=&
-\tfrac12\nabla\cdot f_1
-\nabla\cdot f_5
-\nabla\cdot f_6
-\tfrac12g_{20}
+\tfrac12h_1
+h_9\,.
\label{eq:h5id}
\end{align}

\bibliographystyle{nb}
\bibliography{biblio}{}

\end{document}